\documentclass[10pt,journal,compsoc]{IEEEtran}

\usepackage{cite}
\usepackage{amsmath,amssymb,amsfonts}
\usepackage{algorithmic}
\usepackage{graphicx}
\usepackage{textcomp}
\usepackage{xcolor}
\usepackage{multirow}
\def\BibTeX{{\rm B\kern-.05em{\sc i\kern-.025em b}\kern-.08em
    T\kern-.1667em\lower.7ex\hbox{E}\kern-.125emX}}
\usepackage{array, colortbl}
\usepackage{ulem, comment}
\usepackage{lscape}
\usepackage{longtable}
\usepackage{adjustbox}
\usepackage{slashbox}
\usepackage{csquotes}
\renewcommand{\mkbegdispquote}[2]{\itshape}
\usepackage{soul,color}
\usepackage{tabularray}
\usepackage[justification=centering]{subfig}
\usepackage[many]{tcolorbox}
\usepackage{balance}
\usepackage{mathtools}
\DeclareMathOperator*{\argmax}{arg\,max}

\newcommand{\Foutse}[1]{\textcolor{red}{{\it [Foutse: #1]}}}
\newcommand{\Lionel}[1]{\textcolor{green}{{\it [Lionel: #1]}}}
\newcommand{\etal}{\textit{et al.}}
\newtcolorbox{boxblock}[2][]{
top=0.15in,left=4pt,right=4pt,bottom=4pt,
fonttitle=\bfseries,
colbacktitle=gray,
colback=gray!5,
colframe=gray!40!black,
enhanced,
attach boxed title to top left={xshift=1.5em,yshift=-\tcboxedtitleheight/2},
boxed title style={size=small},
drop shadow={black!50!white},
title=#2,#1}

\begin{document}

\title{Threat Assessment in Machine Learning based Systems
}

\author{Lionel~Nganyewou~Tidjon and Foutse~Khomh\IEEEmembership{, Senior Member,~IEEE}
\IEEEcompsocitemizethanks{\IEEEcompsocthanksitem The authors are with Polytechnique Montréal, Montréal, QC H3C 3A7, Canada. 
E-mail: \{lionel.tidjon, foutse.khomh\}@polymtl.ca
}
}

\IEEEtitleabstractindextext{

\begin{abstract}
Machine learning is a field of artificial intelligence (AI) that is becoming essential for several critical systems, making it a good target for threat actors. Threat actors exploit different Tactics, Techniques, and Procedures (TTPs) against the confidentiality, integrity, and availability of Machine Learning (ML) systems.
During the ML 
cycle, they exploit adversarial TTPs to poison data and fool ML-based systems. In recent years, multiple security practices have been proposed for traditional systems but they are not enough to cope with the nature of ML-based systems. In this paper, we conduct an empirical study of threats reported against ML-based systems with the aim to understand and characterize the nature of ML threats and identify common mitigation strategies. The study is based on 
89 real-world ML attack scenarios from the MITRE's ATLAS database, the AI Incident Database, and the literature; 854 ML repositories 
from the GitHub search and the Python Packaging Advisory database, 
selected based on their reputation.
Attacks from the AI Incident Database and the literature are used to identify vulnerabilities and new types of threats that were not documented in ATLAS. Results show that 
convolutional neural networks were one of the most targeted models among the attack scenarios. ML repositories with the largest vulnerability prominence include TensorFlow, OpenCV, and Notebook.
In this paper, we also report the most frequent vulnerabilities in the studied ML repositories, the most targeted ML phases and models, the most used TTPs in ML phases and attack scenarios. This information is particularly important for red/blue teams to better conduct attacks/defenses, for practitioners to prevent threats during ML development, and for researchers to develop efficient defense mechanisms.
\end{abstract}

\begin{IEEEkeywords}
Machine Learning, Vulnerabilities, Threat Assessment, TTP, Computer security, Artificial Intelligence.
\end{IEEEkeywords}}
\maketitle
\IEEEpeerreviewmaketitle

\begin{figure*}[h]
\centering
\includegraphics[width=1\textwidth]{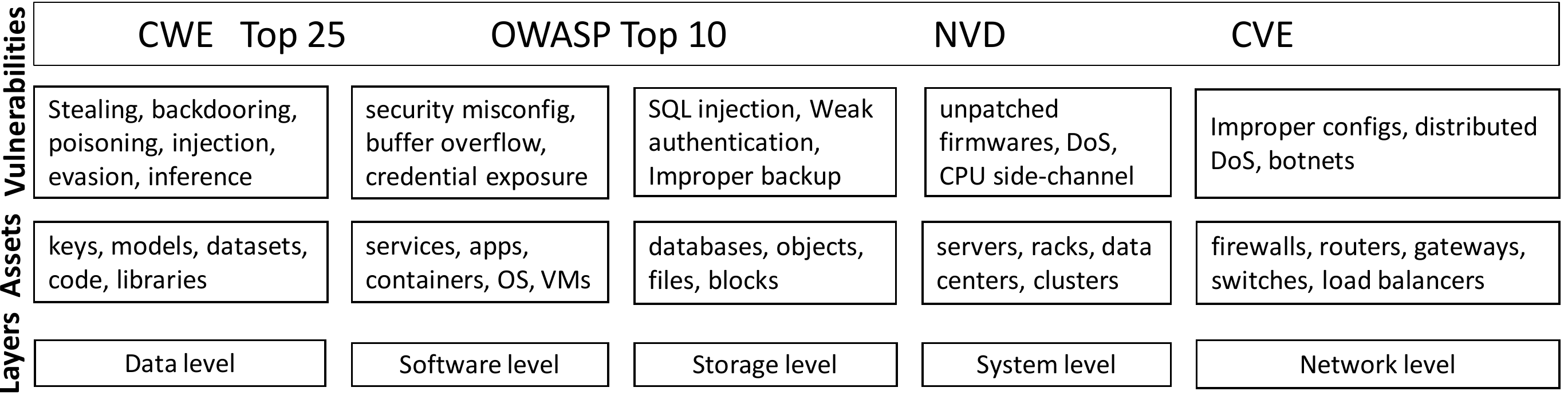}
\caption{Assets and Vulnerabilities}
\label{fig:vuln_arch}
\end{figure*}

\section{Introduction}

Nowadays, Machine Learning (ML) is achieving significant success in dealing with various complex problems in safety-critical domains such as healthcare~\cite{kourou2015machine}
and space~\cite{girimonte2007artificial}. 
ML has also been applied in cybersecurity to detect threatening anomalous behaviors such as spams, malwares, and malicious URLs~\cite{8735821}; allowing a system to respond to real-time inputs, containing both normal and suspicious data, and learn to reject malicious behavior. While ML is strengthening defense systems, it also helps threat actors to improve their tactics, techniques, and procedures (TTPs), and expand their attack surface. Attackers leverage the black-box 
nature of ML models and manipulate input data to affect their performance~\cite{carlini2021extracting, jagielski2018manipulating, biggio2018wild, 8294186}.

Most recent work~\cite{8294186,carlini2021extracting,arp2022and,morris2020textattack,pierazzi2020intriguing,tramer2020adaptive,carlini2019evaluating,abdullah2019practical,eykholt2018robust,biggio2018wild} outlined ML attacks and defenses targeting different phases of the ML lifecycle, i.e., input data, training, inference, and monitoring.
ML-based systems are also often deployed on-premise or on cloud service providers; which increases attack vectors and makes them vulnerable to traditional attacks at different layers: software-level, system-level, and network-level. At the software-level, ML-based systems 
are vulnerable to operating system (OS) attacks since attackers can exploit OS vulnerabilities and bugs to break it.
At the system-level, ML-based systems are vulnerable to several attacks, including CPU side-channel~\cite{7163050} and memory-based~\cite{197211} attacks. At the network-level, ML-based systems can be 
compromised under several attacks~\cite{8735821}, including 
Denial of Service (DoS), botnets, and ransomwares.
To achieve their goals, ML threat actors can poison data and fool ML-based systems using 
\mbox{evasion~\cite{8294186, biggio2018wild, morris2020textattack, abdullah2019practical}}, \mbox{extraction \cite{carlini2021extracting, jagielski2020high, orekondy2019knockoff}}, inference\mbox{~\cite{shokri2017membership, 8294186}}, and poisoning\mbox{~\cite{biggio2018wild, jagielski2018manipulating, morris2020textattack, 8294186, abdullah2019practical}}.
To defend against such threats, adversarial defenses have been proposed including \mbox{~\cite{arp2022and, tramer2020adaptive}}.
Usually, threat TTPs and mitigations are reported in a threat assessment framework to help conduct attack and defense operations. Unfortunately, there is lacking concrete applications of threat assessment in the ML field that provide a broader overview of ML threats, ML tool vulnerabilities, and mitigation solutions.


\textbf{Contributions}. In this work, our goal is to contribute to ML threat assessment by leveraging existing standard frameworks~\mbox{\cite{mitreatlas, mitre_attack}} containing knowledge of TTPs 
as well as the vulnerabilities reported in 
ML repositories. The contributions in this paper include:
\begin{itemize}
    \item new threat TTPs extracted from the AI Incident database~\cite{mcgregor2021preventing} and the literature to complete the MITRE's ATLAS database;
    \item the mapping of threat TTPs from multiple databases to ML phases, models, and tools to evaluate threat impact on ML components;
    \item a vulnerability analysis of ML repositories as well as the dependencies that cause them to foresee potential threats in the use of a particular ML library;
    \item a multi-layer ML mitigation matrix that leverages security guidance from standard databases~\mbox{\cite{attack_desc, mitreatlas, mitre_defend, zero-trust-paper}}, National Institute of Standards and Technology (NIST), and Cloud Security Alliance to help prevent, detect, and respond against ML threats
\end{itemize} 

To achieve this goal, we conduct an empirical study to characterize ML threats, understand their impacts on ML components (i.e., phases, models, tools), and identify common mitigation strategies. 
Results show that deep neural networks (DNNs) such as 
convolutional neural networks (CNNs) 
are one of the most targeted models in attack scenarios. 
ML repositories such as \textit{TensorFlow, OpenCV, and Notebook, Numpy} 
have the largest vulnerability prominence. The most severe dependencies that caused the vulnerabilities include \textit{pickle, joblib, numpy116, python3.9.1,} and \textit{log4j}. \textit{DoS} and \textit{buffer-overflow} were the most frequent in ML repositories.
Our examinations of vulnerabilities and attacks reveal that testing, inference, training,
and data collection are
the most targeted ML phases
by threat actors.
The mitigations of these vulnerabilities and threats include adversarial defenses~\cite{arp2022and, tramer2020adaptive, aigrain2019detecting} and traditional defenses such as software updates, 
and cloud security policies (e.g., zero trust). 
Leveraging our findings, ML read/blue teams can take advantage of the ATLAS TTPs and the newly-identified TTPs from the AI incident database to better conduct attacks/defenses using
the most exploited TTPs and models 
for more impact.
Since ML-based systems are increasingly in production, ML practitioners can leverage these results to prevent vulnerabilities and threats in ML products during their lifecycle. Researchers can also use the results to propose 
theories and algorithms for strengthening defenses.

The rest of this paper is structured as follows.
In Section~\ref{conceptsandrelatedwork}, we define basic concepts such as vulnerabilities, threats; and review the related literature.
Section~\ref{methodology} describes the study methodology for threat assessment while defining some research questions and presents a formal definition of an ML threat. In Section~\ref{evaluation}, we present results while answering to the defined research questions. Section~\ref{mitigation} proposes mitigation solutions for the observed threats and vulnerabilities. 
Section~\ref{threat2valid} discusses threats that could affect the validity of the reported results.
Section~\ref{conclusion} concludes the paper and outlines avenues for future work.

\section{Background and related work}\label{conceptsandrelatedwork}

Before diving in ML threat assessment, generic security concepts such as assets, vulnerabilities, and threats must be defined. This section provides an overview of security concepts and related work.

\subsection{Assets}

In computer security, an asset is any valuable logical or physical object owned by the organization such as data, database, software, hardware, storage, and network devices (see \mbox{Fig~\ref{fig:vuln_arch}}). 
At data level, ML assets can be access keys (tokens, user/password, private/public key, certificates), datasets, models and their parameters, source code, and libraries. At software level, ML assets can be model as a service (MaaS) APIs, ML apps in production, containers and virtual machines (VMs) where ML apps are deployed. At storage level, ML assets can databases, objects (e.g., buckets), files, and blocks to host ML training datasets, models, and code. At system level, ML assets can be servers, racks, data centers, and clusters. At network level, ML assets can be firewalls, routers, gateways, switches, and load balancers.

\subsection{Vulnerabilities}

A vulnerability is a software or hardware flaw that can be exploited by threat actors to execute malicious command and control (C2) operations such as data theft and destruction. 

\subsubsection*{Types of vulnerabilities}

Vulnerabilities occur at different levels: \textit{data}, \textit{software}, \textit{storage}, \textit{system}, and \textit{network} (see \mbox{Fig~\ref{fig:vuln_arch}}). At the \textit{data} level, data assets are vulnerable to model stealing, backdooring~\cite{gao2020backdoor}, poisoning, injection, evasion, and inference. At the \textit{software} level, threat actors look at errors or bugs in ML apps such as buffer overflow, exposed credentials, and security misconfigurations. At the \textit{storage} level, ML databases and cloud storage are vulnerable to weak authentication, improper backup, and SQL injection attacks. At \textit{system} level, they exploit several hardware vulnerabilities including firmware unpatching and CPU side-channel~\cite{7163050} to launch attacks such as DoS,  affecting the ML cloud infrastructure where ML apps are managed. 
At \textit{network} level, threat actors can exploit improper configurations of the network making it vulnerable to distributed DoS and botnet attacks. These vulnerabilities are reported in Common Weaknesses Exposure (CWE) Top 25~\cite{mitrecwe}, OWASP Top 10~\cite{owasp}, National Vulnerability Database (NVD), and Common Vulnerability and Exposures (CVE) standards.   


\subsection{Threats}\label{sota_attack}

A threat exploits a given vulnerability to damage and/or destroy a target asset. Threats can be of two types: \textit{insider} threats and \textit{outsider} threats. \textit{Insider} threats originate from the internal system and they are more often executed by a trusted entity of the system (e.g., employee). \textit{Outsider} threats are operated from the remote/external system.
In the following, we
distinguish between traditional threats and recent machine learning threats. 

\subsubsection*{Traditional threats}\label{trad_attacks}

Adversarial Tactics, Techniques, and Common Knowledge (ATT\&CK)~\cite{attack_desc} is a public and standard knowledge database of attack TTPs. Traditional attack phases are divided into two groups: traditional pre-attack phases and attack phases.

\textit{\underline{Pre-attack}}. The pre-attack phase consists of two tactics: reconnaissance and resource development~\cite{attack_desc}. During reconnaissance, attackers use several techniques including network scanning to find information about a victim such as open ports and OS version (e.g., nmap, censys), 
and use phishing techniques to embed malicious links in emails or SMS messages. During resource development, attackers use several techniques including acquisition of resources to support C2 operations (e.g., domains), 
purchasing a network of compromised systems (botnet) for C2, development of tools (e.g., crawlers, exploit toolkits), 
and phishing.

\textit{\underline{Attack.}} 
Once the pre-attack phase is done, attackers will attempt an initial access to the target victim host or network by delivering a malicious file or links through phishing; exploiting vulnerabilities in websites/softwares used by victims; 
and manipulating software dependencies and development tools prior to their delivery to the final consumer. 
Upon the success of the initial access, 
they will execute malicious code on the victim host/network.
After execution, they will attempt to persist on the target by modifying registries (e.g., Run Keys/Startup Folder), 
and automatically executing at boot.
In addition, attackers will try to gain high-level permissions (e.g., as root/administrator).
To hide their bad actions, they will make sure they are undetected by installed antivirus or Endpoint Detection Response (EDR) tools. 
An attacker can also execute lateral movement techniques such as exploitation of remote services to spread on other hosts or networks for impact. 

\begin{figure*}[h]
\centering
\includegraphics[width=\textwidth]{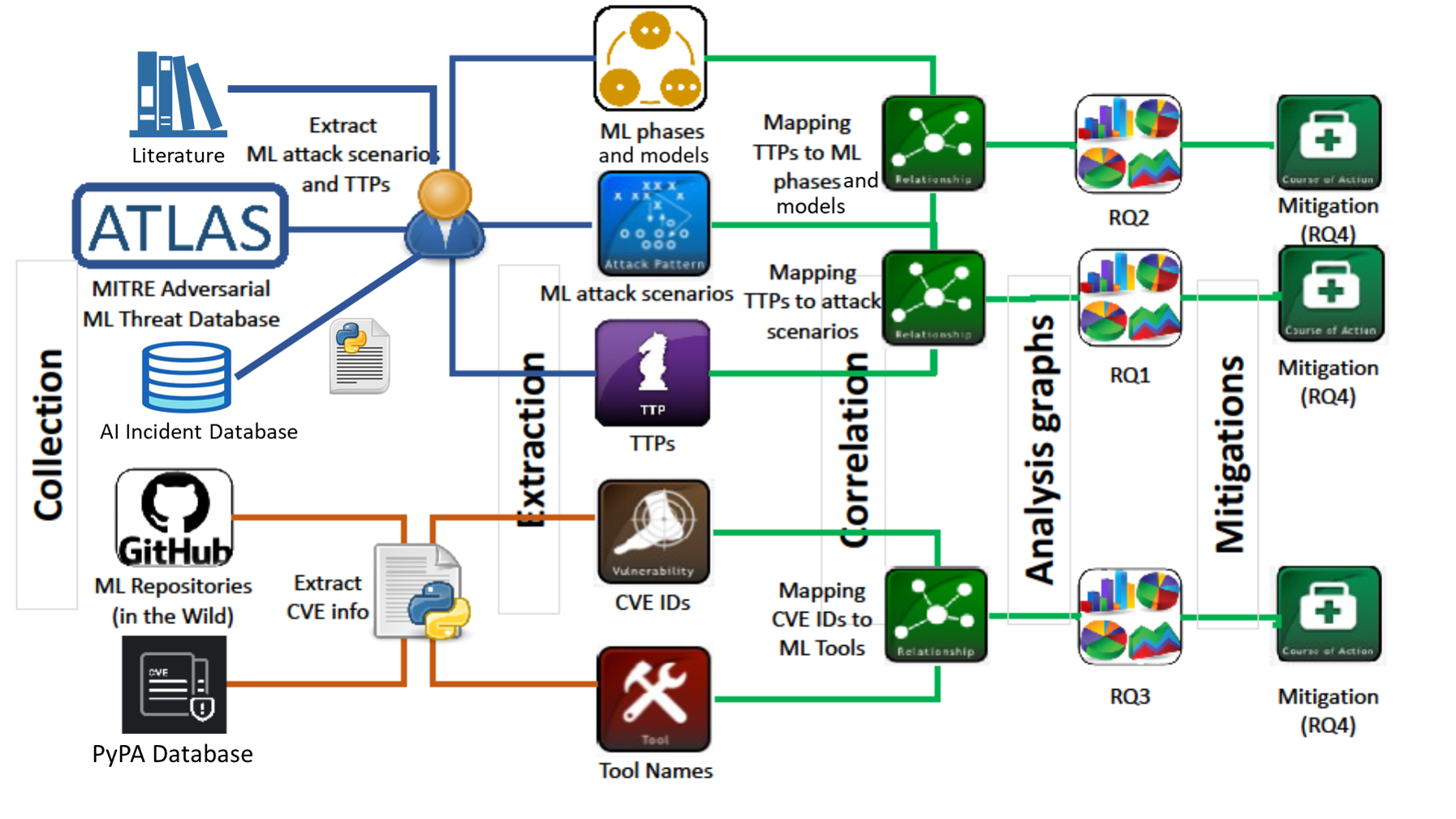}
\caption{Study methodology - Road map}
\label{fig:method1}
\end{figure*}
\subsubsection*{Machine learning threats}

Adversarial Threat Landscape for Artificial Intelligence Systems (ATLAS)~\cite{mitreatlas} is a public and standard knowledge database of adversarial TTPs for ML-based systems~\cite{mitreatlas}.
ML attack phases are divided into two groups: ML pre-attack phases and attack phases.

\textit{\underline{ML Pre-attack.}}
ML pre-attack tactics are similar to the one used in traditional threats, but 
with new techniques and procedures adapted to the ML context~\cite{mitreatlas}. During reconnaissance, Threat actors will search for victim's publicly available research materials such as technical blogs, 
and pre-print repositories; and search for public ML artifacts such as development tools (e.g., TensorFlow). 
For resource development, they will also acquire adversarial ML attack implementations such as adversarial robustness toolbox (ART).

\mbox{\textit{\underline{ML Attack.}}} ML systems are vulnerable to traditional attacks and another kind of attacks that turns their normal behaviors into threatening behaviors called adversarial attacks. Like traditional threats, ML threats target confidentiality, integrity, and availability of data. To achieve the goal, attackers may have full knowledge (white-box), partial knowledge (gray-box), or no knowledge (black-box) of the targeted ML system. In black-box settings, attackers do not have access to the training dataset, model, and the executing code (since assets are hosted in a private corporate network) but they have an access to the public ML API as a legitimated user. This allows them to only perform queries and observe \mbox{outputs~\cite{mitreatlas101}}.

In white-box settings, attackers have a knowledge of the model architecture and can access the training dataset or model to manipulate the training process. In gray-box settings, they have either a partial knowledge of the model architecture or some information about the training process. Whether white-box, gray-box or black attacks, they can be focused on a particular class/sample (targeted) or any class/sample with no specific choice (untargeted) to cause models misclassify inputs. Different techniques are used for attacks: poisoning, evasion, extraction, and inference. During poisoning~\cite{jagielski2018manipulating, biggio2018wild, eykholt2018robust, chen2021badnl, morris2020textattack, abdullah2021hear, abdullah2019practical, abdullah2021sok}, attackers injects false training data to corrupt the learning model (even allowing it to be \mbox{backdoored~\cite{gao2020backdoor}}) to in order to achieve an expected goal at inference time. During evasion~\cite{biggio2013evasion, biggio2018wild, pierazzi2020intriguing, kwon2019selective}, attackers iteratively and carefully modify ML API queries and observing the output at inference \mbox{time~\cite{mitreatlas101}}. The queries seem normal but are misclassified by ML models. During extraction~\cite{carlini2021extracting, jagielski2020high, orekondy2019knockoff, schonherr2018adversarial}, attackers iteratively query the online model~\cite{mitreatlas101} allowing them to extract information about the model. Then, they use this information to gradually train a substitute model that imitates the predictive behaviour of the target model. During inference~\cite{shokri2017membership, choquette2021label, nasr2019comprehensive}, attackers probe the online model with different queries. Based on the results, they can infer whether features are used to train the model or not; allowing them to compromise private data.

The adversarial models used~\cite{8294186} for attack include (1) \textit{fast gradient sign method (FGSM)} which consists in adding noise with the same direction as the gradient of the cost function w.r.t to data, (2) \textit{DeepFool} efficiently computes perturbations that fool deep networks, (3) \textit{Carlini and Wagner (C\&W)} is a set of three attacks against defensive distilled neural networks~\cite{papernot2016distillation}, 
(4) \textit{Jacobian-based saliency map (JSMA)} saturates a few pixels in an image to their maximum/minimum values, (5) \textit{universal adversarial perturbations} are agnostic-image perturbations that can fool a network on \textit{any} image with high probability, (6) \textit{Basic Iterative Method (BIM)} is an iterative version of the FGSM, (7) \textit{one pixel} is when a single pixel in the image is changed to fool classifiers, (8) \textit{Iterative Least-likely Class Method (ILCM)} is an extension of BIM where an image label is replaced by a target label of the least-likely class predicted by a classifier, and (9) \textit{Adversarial Transformation Networks (ATNs)} turns any input into an adversarial attack on the target network, while disrupting the original inputs and outputs of the target network as little as possible. 



\subsection{Related work}

In computer security, threat assessment is a continuous process that consists in identifying, analyzing, evaluating, and mitigating threats. It has been applied for assessing threats in traditional systems~\cite{4652578} but to date, there are no concrete applications of this process for ML-based systems. The ATLAS framework is the first thorough real-world attempt of 
ML threat assessment built by the MITRE Corporation jointly with some organisations including Microsoft and Palo Alto Networks~\cite{mitreatlas}. Most recent work used ATT\&CK for threat assessment. 
\mbox{Kumar \etal{}~\cite{9283867}} 
identified gaps during ML development, deployment, and when ML-based system is under attack. Then, 
they enumerated ML security aspects (e.g., static/dynamic analysis of ML systems, auditing and logging) to be considered in industry. The mitigations proposed in this paper include ML security aspects by \mbox{Kumar \etal{}} with additional security aspects in standards (ATT\&CK mitigations, Cloud Security Alliance, NIST), and we better organize it in terms of layer (data level to cloud level, assets, and ML pipelines) and in terms of course of action (i.e., harden, detect, isolate, evict) following MITRE D3FEND\mbox{~\cite{mitre_defend}}.
Lakhdhar \etal{}~\cite{9474289} 
mapped new discovered vulnerabilities to potential attack tactics in ATT\&CK by extracting features (e.g., CVSS severity score, vendor) to train a RandomForest-based model. 
This approach is limited to the ATT\&CK database and maps only vulnerabilities to ATT\&CK TTPs. 
Kuppa \etal{}~\cite{10.1145/3465481.3465758} also mapped  CVE's to ATT\&CK techniques using a multi-head joint embedding neural network, trained using threat reports from Symantec and Google Project Zero.
Like~\cite{9474289}, this approach is limited to the ATT\&CK database and maps only vulnerabilities to ATT\&CK TTPs.

Our threat assessment leverages ATLAS, ATT\&CK and other databases like the AI Incident Database~\cite{mcgregor2021preventing} to have a complete set of TTP definitions for ML threat characterization, ML phase and model mapping. The threat assessment process is also coupled with a vulnerability assessment that identifies, analyzes, evaluates, and mitigates vulnerabilities in ML-based systems. In addition, it leverages defenses frameworks (ATT\&CK mitigations~\cite{mitre_attack_mitigation}, D3FEND~\cite{mitre_defend}), Cloud Security Alliance~\cite{csa_report} and NIST security guidelines~\cite{mccarthy1800identity, hu2020general, souppaya2013guide, scarfone2008technical, scarfone2008sp, karygiannis2002wireless, cooper2018security} to provide mitigations during the ML lifecycle. This combination provides a complete and broader end-to-end threat assessment of ML assets. In the following section, we show how
ATLAS TTPs affect 
ML components at different phases of the ML lifecycle and how traditional threats/vulnerabilities affect ML repositories.

\section{Study Methodology}~\label{methodology}
The \textbf{goal} of this study is to analyze ML threat behaviors (i.e., common entry points, prominent threat tactics, common TTP stages)  and understand their impact on ML components (i.e., vulnerable ML phases, models, tools and, related-dependencies) 
by leveraging ML threat knowledge in ATLAS, ATT\&CK and the AI Incident Database as well as vulnerabilities from  ML repositories to foresee potential threats in the use of a particular package or library.
The \textbf{perspective} of this study is that of ML red/blue teams, researchers, and practitioners looking to 
gain knowledge about ML threat TTPs, prevent and defend against those threats, build, and deploy secure ML products from staging to production environment. The \textbf{context} of this study consists of 89 real-world ML attack scenarios (15 from ATLAS, 60 from the AI Incident Database, 13 from the literature) and 854 ML repositories (845 from GitHub, 9 from PyPA\mbox{~\cite{pypa-db}}). In Fig~\ref{fig:method1}, the road map of the study is shown.   
To achieve our goal, we address the following research questions:
\begin{quote}
	\textbf{RQ$_1$:} \textit{What are the prominence and common entry points of threat TTPs exploited in ML attack scenarios?} This research question aims to gain knowledge about ML threat TTPs in order to allow the implementation of better 
	defenses 
	strategies. In this question, we also examine the 
	execution flows of ML attack scenarios to identify the most used TTPs and TTP sequences adopted by attack scenarios, in order to defend against them.
\end{quote}

\begin{quote}
	\textbf{RQ$_2$:} \textit{What is the effect 
	of threat TTPs on different ML phases and ML models?} This research question aims to understand the effect of 
	each threat tactic identified in \textbf{RQ$_1$} on ML phases and ML models, 
	with the aim to help secure each component of the ML lifecycle.
	Through this research question, we also aim to identify 
	the most targeted ML phases as well as the most used threat TTPs in the different ML phases.
\end{quote}

\begin{quote}
	\textbf{RQ$_3$:} \textit{Are there new real-world threats in the AI Incident Database, the literature, and ML repositories that were not documented in the ATLAS database?} This research question aims to assess the completeness of ATLAS with the aim to identify other security threats that may have been overlooked as well as the most vulnerable ML repositories, the dependencies that cause them, and the most frequent vulnerabilities in the ML repositories. 
\end{quote}

\begin{quote}
	\textbf{RQ$_4$:} \textit{How can ML stakeholders harden their ML assets and prevent ML threats during development to production stage from data level to cloud level?} This research question aims to provide end-to-end mitigation solutions to ensure security of ML assets based on ~\mbox{\cite{attack_desc, mitreatlas, mitre_defend, csa_report}}. 
\end{quote}

\begin{quote}
\end{quote}

\subsection{Threat Model}\label{threat_model}

\textbf{Goal.} Attackers aim to affect the confidentiality, integrity, and availability of data (e.g., training data, features, model) depending of threat objectives. Poisoning attacks can affect data integrity. Extraction attacks can allow one to steal models or features; thus affecting confidentiality.  

Let $a \in A$ be an asset, where $A$ is a set of assets from the system $\mathcal{S}$. An asset $a$ can be owned or accessed by an entity (e.g., user, user group, program, set of program), denoted $E$. $\mathcal{E}$ denotes the set of all entities and $E \in \mathcal{E}$. Let $AC_{\mathcal{S}}: A \times \mathcal{E} \rightarrow R$ be a function, that defines the level of privilege that an entity $E$ has on an asset $a$ or an asset group $A_g \subseteq A$, under the system $\mathcal{S}$. $R$ is a set of right access and it can take values (1) $R=\{\textsf{none}, \textsf{user}, \textsf{root}\}$ meaning that entities can have either no privilege ($\textsf{none}$), user access on $A$ ($\textsf{user}$), and full access on $A$ ($\textsf{root}$); or (2) $R=\{\textsf{none}, \textsf{read}, \textsf{write}\}$ meaning that entities can have no privilege ($\textsf{none}$), read access on $A$ ($\textsf{read}$), and write access on $A$ ($\textsf{write}$). When $AC_{AWS}('model.pkl', 'ml\_api') = \textsf{root}$, it means that the Amazon Web Service (AWS) ML API service $ml\_api$ have full access on the pickled model file $model.pkl$. When $AC_{VM}('training.csv', 'john') = \textsf{write}$, it means that user $john$ can modify or delete the training data file $training.csv$ in th virtual machine $VM$.

Let $P_1,...P_n$ be a set of premises and $C$ the goal to achieve. This relation is represented by
\[
   \frac{P_1,...P_n}{C}
\]
It also means that $C$ can succeed when properties $P_1,...P_n$ are satisfied. Based on \cite{kindred1999theory}, we define the following notations. The notation
\[
   \xmapsto[]{\text{a}}E
\]
means that $a$ is $E$'s asset. Given $k_E \in K$ a protection property (e.g., encryption key, certificate, token), the notation 
\[
   \{a\}_{k_E}
\]
means that the protection $k_E$ is enforced on asset $a$ by an entity $E$. Let $E1, E2 \in \mathcal{E}$ be two entities that share an asset $a$. The notation
\[
   E1 \xleftrightarrow[]{\text{a}} E2
\]
means that $a$ is shared by $E1$ and $E2$. The sharing is satisfied when $E1$ send $a$ to $E2$ and $E2$ send $a$ to $E1$ as follows,
\[
   \frac{E1 \xrightarrow[]{\text{a}} E2,E2 \xrightarrow[]{\text{a}} E1}{E1 \xleftrightarrow[]{\text{a}} E2}
\]
Let $m: A \rightarrow C$ be a model function that takes data in $A$ and return decisions in $C$ based on the inputs. $C$ can be two classes (i.e., $\{c_1, c_2\}$) or multiple classes (i.e., $\{c_1, c_2, ...,c_n\}$), where $c_1, c_2, ...,c_n \in C$.

\textbf{Knowledge.} In black-box settings, an attacker $AT$ does not have direct access~\cite{mitreatlas101} to ML assets $A_V$ of the target victim $V$ (i.e., model, executing code, datasets), i.e.,  $\forall a_V \in A_V, AC_V(a_V, AT) = none$. They have only access to the ML inference API using an access token $k_V$ obtained as a legitimate user from the victim's platform $V$, i.e.,

\[
   \xmapsto[]{\{a_{AT}\}} AT
\]
where $a_{AT} \in A$ are data crafted offline by attacker $AT$ to be send via API. During the attack, $AT$ performs queries using the victim's ML inference API and observe outputs. To do so, $AT$ sends an online request with the crafted data $a_{AT}$ using access token $k_{V}$, i.e., 

\[
   AT \xrightarrow[]{\{a_{AT}\}_{k_{V}}} V
\]

Then, $AT$ will receive prediction responses and analyze it to further improve its data for attack, i.e.,

\[
   V \xrightarrow[]{\{m_V(a_{AT})\}} AT
\]

where $m_V$ is the executed model behind the ML inference API of the target victim $V$. 

In white-box settings, attacker $AT$ may have an internal access to some ML assets $\tilde{A}_V \subseteq A_V$ of the target victim $V$ (e.g., model, training data), i.e.,  
\[
\forall a_V \in \tilde{A}_V, AC_V(a_V, AT) \in \{\textsf{read},\textsf{write}\}
\] 

Then, $AT$ can perform several state-of-the-art attack techniques such as poisoning, evasion, extraction, and inference (see Section~\ref{sota_attack}).

\textbf{Specificity.} In adversarial settings, ML threats can target a specific class/sample for misclassification (adversarially targeted) or any class/sample for misclassification (adversarially untargeted). The goal of $AT$ is to maximize the loss $L$ so that model $m_V$ misclassifies input data,

\[
   \argmax_{a} L(m_V(a), c)
\]

where $a \in A$ is an input data, $c \in C$ is a target class, and $m_V(a)$ is the predicted target data given $a$. To achieve the goal, $AT$ can execute a targeted or untargeted attack affecting integrity and confidentiality of data~\cite{barreno2010security, jagielski2018manipulating}. 

When attack is targeted, $AT$ substitutes the predicted class $c$ by adding a small pertubation $\theta_u(a, c)$ so that 
\[
m_V(a_{AT}) = c
\]
, where $a_{AT} = a + \theta_u(a, c)$ is an adversarial sample. 

In untargeted attack, $AT$ adds a small pertubation $\theta_t(a)$ to input $a$ so that 
\[
m_V(a_{AT}) \neq m_V(a)
\]
, where $a_{AT} = a + \theta_t(a)$ is an adversarial sample. ML threats can also leverage traditional TTPs to achieve goal. 

In traditional settings, threat actors can actively pursue and compromise the target system while maintaining anonymity (traditionally targeted) or can simply spread on the network as possible without a particular target (traditionally targeted). Terms \textit{Adversarially} (resp. \textit{Traditionally}) are used to distinguish the attack specificity in adversarial settings (resp. traditional settings). In traditional attacks, $AT$ targets assets such as user accounts, servers, virtual machines, databases, and networks by bypassing authentication and firewalls. This allows him to get full control of the ML assets of $V$, i.e., $\forall a_V \in A_V, AC_V(a_V, AT) = \textsf{root}$ and cause more damages.

\textbf{Capability.} To launch ML attacks, threat actors use the following tactics\mbox{~\cite{mitreatlas}}: Reconnaissance, Resource Development, Initial Access, ML Model Access, Execution, Persistence, Defense Evasion, Discovery, Collection, ML attack staging, Exfiltration, and Impact. During Reconnaissance and Resource Development, attackers gather information (e.g., papers, repositories) and setup C2 infrastructure to start the attack. During initial access, they try to gain access to the victim infrastructure containing ML artifacts. Then, they will try to gain access to the internals of the model and the physical environment (ML Model Access). Attackers can also run remote-controlled malicious code to steal data (Execution). To persist on the ML system, they can use backdoored ML models or keep access to the target (Persistence). Attackers also leverage evasion tactics~\cite{biggio2013evasion, biggio2018wild, pierazzi2020intriguing, kwon2019selective} to bypass classifiers (Evasion). When attack succeeded, they can collect data for exfiltration. During attack staging, they can train proxy models, craft adversarial data, and poison the target model for impact (e.g., human loss, ML system destruction).

\subsection{Data collection}

To select data sources, we define three criteria: more recent data, consistency, and reputation. 
The first criteria ensures that data sources contain recent information about ML vulnerabilities and threats. The ATLAS database has been selected since it is recent, i.e., containing threats from 2019 to 2021~\cite{mitreatlas}. The well-known ATT\&CK database also contains enterprise attacks 
from 2018 to 2021 and it is used to complete TTP definitions in ATLAS; since some attack scenarios employed both tactics from ATLAS and ATT\&CK. In addition, the AI Incident database contains recent real-world AI incident information from 2003 to 2022\mbox{~\cite{mcgregor2021preventing}}. Consistency criteria checks whether data sources have a significant amount of vulnerability or threat information. ATLAS contains 15 ML attack scenarios, 12 tactics, and 30 techniques with 43 sub-techniques~\cite{mitreatlas}. Since ATLAS is not consistent, 60 real-world threats are collected from the AI Incident database\mbox{~\cite{mcgregor2021preventing}} between 2018 to 2022 and 14 threats from the literature between 2010 to 2021. Similar to ATLAS, ATT\&CK contains 14 enterprise tactics and 188 enterprise techniques with 379 sub-techniques~\cite{mitre_attack}. Reputation criteria selects the best data sources based on the reputation of the maintaining organizations, the number of stars for GitHub repositories, and the frequency 
of updates. ATLAS and ATT\&CK are maintained by top organisations such as the MITRE corporation, Microsoft, McAfee, Palo Alto Networks, IBM, and NVIDIA. The AI Incident Database contains real-world attack information from top medias such as Forbes, BBC, New York Times, and CNN. Most papers~\cite{carlini2021extracting,biggio2013evasion,barreno2010security,carlini2017adversarial,wallace2020imitation,abdullah2021sok,chen2021badnl,choquette2021label,papernot2016transferability,goodfellow2014generative,papernot2017practical,cisse2017parseval,athalye2018obfuscated,jagielski2020high} selected in the literature are also well-known in the ML community. 

ML repositories from the GitHub search are also selected using the reputation criteria such as the number of stars and the notoriety of the maintaining organizations.
To select reputed repositories, we search 
Github code search API using the following query :  \textit{machine-learning in:topic stars:$>$1000 sort:stars}. This query returns repositories with a number of stars greater than 1000, 
sorted in descending order. ML repositories with at least 1000 stars are commonly considered as reputed or promising~\cite{LF-hosting-project}.
This search returned 845 repositories.
This corresponds to a confidence level of 99\% with an error margin of 4.44\%. We have also picked 9 ML repositories from the PyPA database~\cite{pypa-db}: notebook, jupyter-server, jupyterhub, jupyterlab, pyspark, pandas, numpy, nltk, and protobuf. In fact, the PyPA database contains only 14 ML repositories among the 425 available repositories, including 5 ML repositories that were already retrieved by our search query.
The statistical description of the 845 selected projects can be found in our replication package~\cite{tech-report-v1}.



\begin{figure}[h]
\centering
\includegraphics[width=0.48\textwidth]{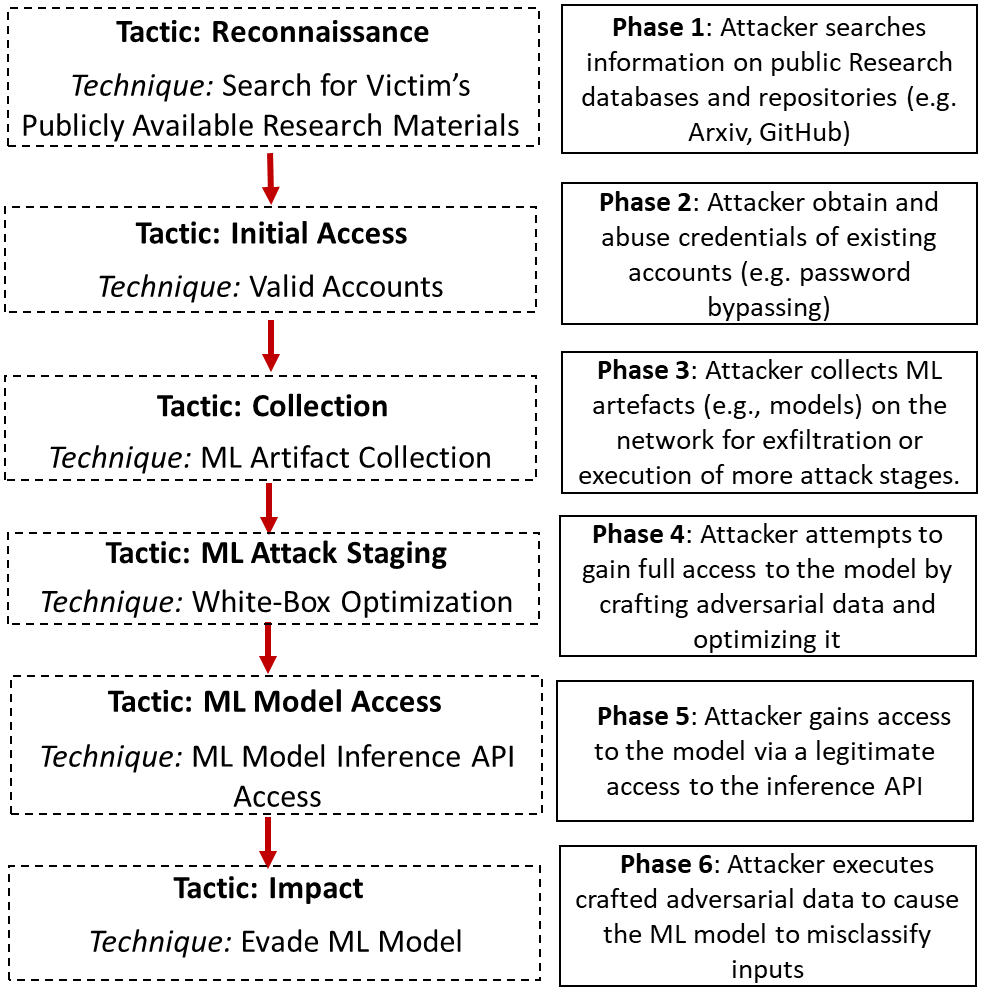}
\caption{TTP definitions for \textit{Microsoft-Azure service} attack}
\label{attack-example}
\end{figure}
\subsection{Data processing}

Data processing consists in three phases: extraction, correlation, and metric computation.
Threat information is extracted manually from the ATLAS database and the literature by reading ML attack scenarios, looking at the tactics used, and finding the tactic descriptions referred in ATLAS or ATT\&CK. The AI Incident database has been also scrutinized to get recent ML real-world attacks using a crawler\mbox{~\cite{tech-report-v1}} and keywords such as \textit{attack}. Next, vulnerability and threat information is extracted from ML repositories using filtering. Filtering is done by searching 
query keywords: 
\textit{vuln} for vulnerability, \textit{attack} for attack and attacker, \textit{threat} for threat, 
\textit{secur} for secure and security, and \textit{cve} for CVE code. These signatures are commonly used by cybersecurity teams to describe potential security issues, threats or vulnerabilities.


\subsubsection{Data extraction}\label{dextract}

We extract data from the ATLAS database, the AI Incident database, the literature, and ML repositories. 

\textbf{ATLAS database.} We extract all the 15 ML attack scenarios provided as well as their TTP definitions referred in ATLAS (12 tactics, 30 techniques, 43 sub-techniques) and ATT\&CK (14 enterprise tactics, 188 enterprise techniques, 379 sub-techniques) by reading the description of each ML attack scenario, finding ATLAS and ATT\&CK descriptions of the tactics
used during the stages of the ML attack scenario. We also extract the attack goal, knowledge, and specificity for each attack scenario following the threat model (see Section \ref{threat_model}). In~\cite{mitreatlascase}, an attack scenario is decomposed into phases which are further decomposed into steps following ML attack phases described in the threat model. Each attack phase represents a tactic and the associated execution steps are techniques.

An attack scenario is illustrated in Fig.~\ref{attack-example}. The attack called \textit{Microsoft - Azure Service}~\cite{mitreatlascase} was executed by the Microsoft Azure Red Team and Azure Trustworthy ML Team against the Azure ML service in production. It has 6 phases. In Phase 1, they collected information necessary for the attack such as Microsoft publications on ML model architectures and open source models. Next, the team used valid accounts to access the internal network (Phase 2). Then, they found training data and model file of the target ML model that allowed them to execute further ML attack stages (Phase 3). During ML attack staging, they crafted adversarial data using target data and model (Phase 4). In Phase 5, they exploited an exposed inference API to gain legitimate access to the Azure ML service. Then, they successfully executed crafted adversarial data on the online ML service. This attack targets the confidentiality (unauthorized model/training data access), integrity (poisoning by crafting adversarial data), and the availability (causing disruption of the ML service) of the ML system. The attack knowledge is white-box since attackers have full access to the training data and the model. The attack specificity is adversarially untargeted since threat actors do not target a specific class of the ML model.



\begin{table}[]
\caption{\label{add-attacks} A sample of the ML real-world attacks from the AI Incident database}
\centering
\resizebox{\columnwidth}{!}{
\begin{tabular}{|l|l|l|l|}
\hline
\textbf{Title}                                                                                                                                                          & \textbf{Description}                                                                                                                                                                                & \textbf{Date} & \textbf{References}                                                                                                                                                                     \\ \hline
\begin{tabular}[c]{@{}l@{}}Indian Tek Fog \\ Shrouds an \\ Escalating \\ Political War\end{tabular}                                                                     & \begin{tabular}[c]{@{}l@{}}Technology is being\\  used against   the \\ people, and those \\ in the world's largest \\ democracy are the \\ latest victims....\end{tabular}                         & 2022          & \begin{tabular}[c]{@{}l@{}}https://www.bloomberg\\ .com/opinion/articles/\\ 2022-01-12/india-s-tek\\ -fog-shrouds-an-escal\\ ating-political-war-\\ against-modi-s-critics\end{tabular} \\ \hline
\begin{tabular}[c]{@{}l@{}}Meta says it's \\ shut down \\ a pro-Russian \\ disinformation \\ network, \\ warns of a   \\ social media\\  hacking operation\end{tabular} & \begin{tabular}[c]{@{}l@{}}Facebook (FB)'s \\ parent Meta said \\  Monday it has caught \\ dozens of fake, \\ pro-Russian \\ accounts, groups and\\  pages across its \\ platforms....\end{tabular} & 2022          & \begin{tabular}[c]{@{}l@{}}https://edition.cnn.com/\\ 2022/02/28/tech/meta\\ -russia-ukraine\\ -disinformation\\ -network/index.html\end{tabular}                                       \\ \hline
\begin{tabular}[c]{@{}l@{}}Libyan Fighters \\ Attacked by a \\ Potentially \\ Unaided\\  Drone, \\ UN Says\end{tabular}                                                 & \begin{tabular}[c]{@{}l@{}}A military drone that \\ attacked soldiers \\ during a battle in \\ Libyia's civil war \\ last year may have\\ done so without \\ human control,....\end{tabular}        & 2021          & \begin{tabular}[c]{@{}l@{}}https://www.nytimes\\ .com/2021/06/03/\\ world/africa/libya\\ -drone.html\end{tabular}                                                                       \\ \hline
\end{tabular}
}
\end{table}

\begin{table}[]
\centering
\caption{ML real-world attack TTPs from the AI Incident database}
\label{new_att}
\resizebox{\columnwidth}{!}{
\begin{tabular}{|l|l|}
\hline
\textbf{Title}                                                                                                                                                          & \textbf{Tactics and Techniques}                                                                                                                                                                                                                                             \\ \hline
\begin{tabular}[c]{@{}l@{}}India's Tek Fog Shrouds \\an Escalating Political War\end{tabular}                                                               & \begin{tabular}[c]{@{}l@{}}\textit{\textbf{Resource Development}}(Establish Accounts),\\ \textit{\textbf{Initial Access}} (Valid Accounts), \\\textit{\textbf{ML Attack Staging}} (Create Proxy ML model: \\Use Pre-Trained Model),\\\textit{\textbf{Exfiltration}} (Exfiltration via Cyber Means)\end{tabular} \\ \hline
\begin{tabular}[c]{@{}l@{}}Meta says it's shut down a \\pro-Russian disinformation \\network, warns of a social\\ hacking operation\end{tabular} & \begin{tabular}[c]{@{}l@{}}\textit{\textbf{Resource Development}} (Establish Accounts),\\\textit{\textbf{Initial Access}} (Valid Accounts),\\ \textit{\textbf{Exfiltration}}  (Exfiltration via Cyber Means)\end{tabular}                                                                           \\ \hline
\begin{tabular}[c]{@{}l@{}}Libyan Fighters Attacked by \\ a Potentially Unaided Drone, \\UN Says\end{tabular}                                                 & \begin{tabular}[c]{@{}l@{}}\textit{\textbf{Reconnaissance}} (Active Scanning),\\ \textit{\textbf{Impact}} (Cost Harvesting)\end{tabular}                                                                                                                                          \\ \hline
\begin{tabular}[c]{@{}l@{}}Fraudsters Cloned Company\\Director's Voice In \$35M \\ Bank Heist, Police Find\end{tabular}                         & \begin{tabular}[c]{@{}l@{}}\textit{\textbf{ML Attack Staging}} (Create Proxy ML Model: \\Use Pre-trained Model),\\ \textit{\textbf{Impact}} (Cost Harvesting)\end{tabular}                                                                                                                                                                                                                        \\ \hline
\begin{tabular}[c]{@{}l@{}}Poachers evade KZN park's\\ high-tech security and kill \\four rhinos for their horns\end{tabular}                          & \begin{tabular}[c]{@{}l@{}}\textit{\textbf{Defense Evasion}} (Evade ML Model),\\ \textit{\textbf{Impact}} (Evade ML Model)\end{tabular}                                                                                                                                                                     \\ \hline
\begin{tabular}[c]{@{}l@{}}Tencent Keen Security Lab: \\Experimental Security \\Research of Tesla Autopilot\end{tabular}                                   & \begin{tabular}[c]{@{}l@{}}\textit{\textbf{ML Attack Staging}} (Craft Adversarial Data),\\ \textit{\textbf{Impact}} (Evade ML Model, Cost Harvesting)\end{tabular}                                                                                                                                                         \\ \hline
\begin{tabular}[c]{@{}l@{}}Three Small Stickers in \\Intersection Can Cause Tesla \\Autopilot to Swerve Into \\Wrong Lane\end{tabular}                   & \begin{tabular}[c]{@{}l@{}}\textit{\textbf{ML Attack Staging}} (Craft Adversarial Data),\\ \textit{\textbf{Impact}} (Evade ML Model, Cost Harvesting)\end{tabular}                                                  \\ \hline
\begin{tabular}[c]{@{}l@{}}The DAO Hack - Stolen\\ \$50M \& The Hard Fork.\end{tabular}                                                                        & \begin{tabular}[c]{@{}l@{}}\textit{\textbf{Impact}} (Cost Harvesting, Denial of Service)\end{tabular}                                                                                                                                                          \\ \hline
\end{tabular}
}
\end{table}

\textbf{AI Incident database.} We automatically extract 64 recent ML real-world attacks using query \textit{attack} and the period between 2018 and 2022 using a crawler\mbox{~\cite{tech-report-v1}}. Then, we remove row duplicates and it remains 60 ML real-world attacks. Then, we manually read the description of these attacks using the reference link to identify if there are potential TTPs similar to those in ATLAS/ATT\&CK and explicit mention of the models used, the attack goal, the attack knowledge and specificity. A sample of the extracted attacks (title, description, date, references) is shown in Table\mbox{~\ref{add-attacks}}. The remainder can be found in the replication package\mbox{~\cite{tech-report-v1}}. Among the 60 new threats\mbox{~\cite{tech-report-v1}}, only 13 ML threats have some threat tactics/techniques mentioned in  ATLAS/ATT\&CK. Some records have different names but was similar; thus we have considered 8 unique records. Table\mbox{~\ref{new_att}} shows a description of the new TTPs extracted in ML threats in the form \textit{\textbf{atlas\_tactic}} (atlas\_tech1, atlas\_tech2,...), where \textit{\textbf{atlas\_tactic}} (resp. atlas\_tech) is an ATLAS tactic (resp. technique). These 8 ML real-world attacks were not documented in ATLAS and are used to complete case studies in the ATLAS database.

\begin{figure}[h]
\centering
\includegraphics[width=0.48\textwidth]{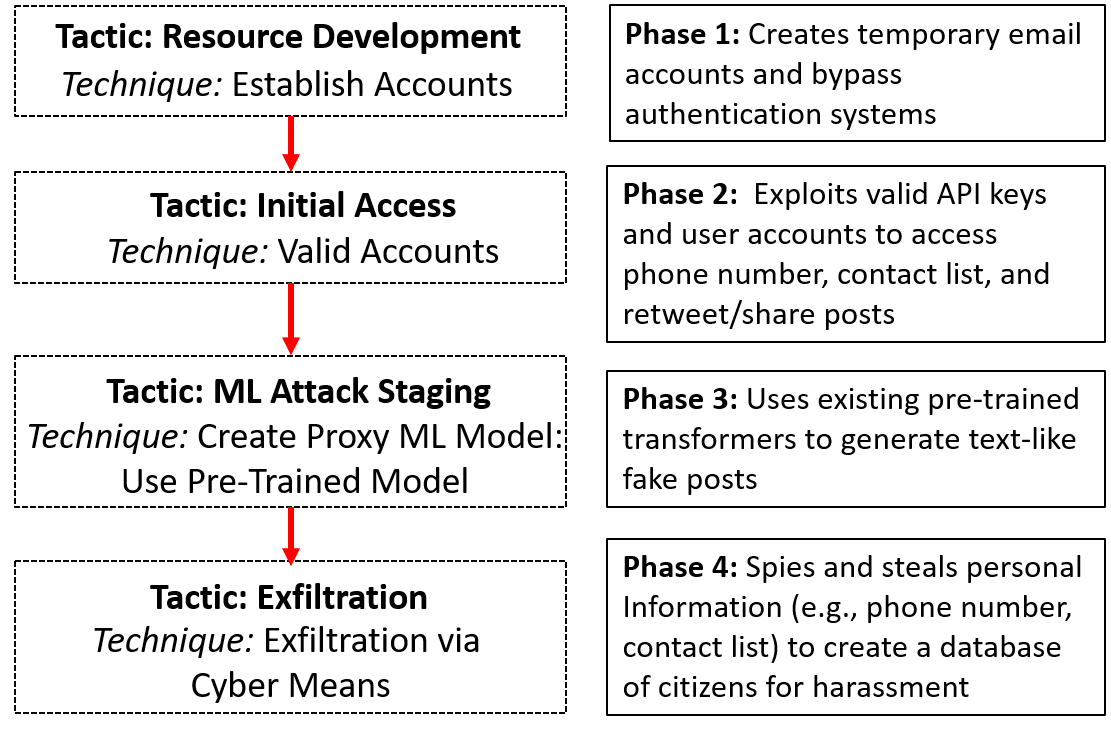}
\caption{TTP definitions for new \textit{Indian Tek Fog Shrouds an Escalating Political War} attack}
\label{attack-example1}
\end{figure}

In Table\mbox{~\ref{new_att}}, let consider the ML real-world attack \textit{Indian Tek Fog Shrouds an Escalating Political War}. Following the reference link in Table~\ref{add-attacks}, Tek Fog is an ML-based bot app used to distort public opinion by creating temporary email addresses and bypassing authentication systems (i.e., WhatsApp, Facebook, Instagram, Twitter and Telegram) to send fake news, automatically hijacking Twitter and Facebook trends such as retweeting/sharing posts to amplify propaganda, phishing inactive WhatsApp accounts and spying personal information (e.g., phone number, contact list), and building a database of citizens for harassment. The bot may uses Transformer model such as GPT-2 to generate coherent text-like messages~\cite{thewire_attack}. This allows to deduce the 4 following ML TTPs (see Fig.~\ref{attack-example1}): Resource Development (Establish Accounts), Initial Access (Valid Accounts), ML Attack Staging (Create Proxy ML Model: Use Pre-Trained Model), and Exfiltration (Exfiltration via Cyber Means). This attack targets the confidentiality (bypass authentication systems, spy personal information) and integrity (post fake news about the victim) of the ML system. The attack specificity is traditionally targeted; since threat actors target specific inactive WhatsApp accounts to spy personal information. There is no detail about the attack knowledge in the adversarial context.

\begin{table}[]
\centering
\caption{ML real-world attack TTPs from the literature}
\label{new_att_lit}
\resizebox{\columnwidth}{!}{
\begin{tabular}{|l|l|}
\hline
\textbf{Title}                                                                                                                                                          & \textbf{Tactics and Techniques}                                                                                                                                                                                                                                             \\ \hline
\begin{tabular}[c]{@{}l@{}}Carlini \etal{} \cite{carlini2021extracting}\end{tabular}                                                               & \begin{tabular}[c]{@{}l@{}}\textit{\textbf{ML Attack Staging}} (Create Proxy ML model: \\Use Pre-Trained Model),\\\textit{\textbf{Exfiltration}} (Exfiltration via ML Inference\\ API: Extract ML Model)\end{tabular} \\ \hline
\begin{tabular}[c]{@{}l@{}}Biggio \etal{} \cite{biggio2013evasion}\end{tabular} & \begin{tabular}[c]{@{}l@{}}\textit{\textbf{ML Attack Staging}} (Craft Adversarial Data),\\\textit{\textbf{Defense Evasion}} (Evade ML Model),\\\textit{\textbf{Impact}} (Evade ML Model)\end{tabular}                                                                           \\ \hline
\begin{tabular}[c]{@{}l@{}}Barreno \etal{} \cite{barreno2010security} \end{tabular}                                                 & \begin{tabular}[c]{@{}l@{}}\textit{\textbf{ML Attack Staging}} (Craft Adversarial Data),\\\textit{\textbf{Defense Evasion}} (Evade ML Model),\\\textit{\textbf{Impact}} (Evade ML Model)\end{tabular}                                                                                                                                          \\ \hline
\begin{tabular}[c]{@{}l@{}}Carlini \etal{} \cite{carlini2017adversarial}\end{tabular}                         & \begin{tabular}[c]{@{}l@{}}\textit{\textbf{ML Attack Staging}} (Craft Adversarial data),\\\textit{\textbf{Defense Evasion}} (Evade ML Model),\\\textit{\textbf{Impact}} (Evade ML Model)\end{tabular}                                                                                                                                                                                                                        \\ \hline
\begin{tabular}[c]{@{}l@{}}Wallace \etal{} \cite{wallace2020imitation} \end{tabular}                          & \begin{tabular}[c]{@{}l@{}}\textit{\textbf{ML Attack Staging}} (Create Proxy ML model: \\Use Pre-Trained Model),\\\textit{\textbf{Exfiltration}} (Exfiltration via ML Inference\\ API: Extract ML Model)\end{tabular}                                                                                                                                                                    \\ \hline
\begin{tabular}[c]{@{}l@{}}Abdullah \etal{} \cite{abdullah2021sok}\end{tabular}                                   & \begin{tabular}[c]{@{}l@{}}\textit{\textbf{ML Attack Staging}} (Craft Adversarial Data),\\\textit{\textbf{Defense Evasion}} (Evade ML Model),\\\textit{\textbf{Impact}} (Evade ML Model)\end{tabular}                                                                                                                                                         \\ \hline
\begin{tabular}[c]{@{}l@{}}Chen \etal{} \cite{chen2021badnl}\end{tabular}                   & \begin{tabular}[c]{@{}l@{}}\textit{\textbf{ML Attack Staging}} (Craft Adversarial Data:\\ Insert Backdoor Trigger),\\ \textit{\textbf{Persistence}} (Backdoor ML Model: \\Inject Payload)\end{tabular}                                                  \\ \hline
\begin{tabular}[c]{@{}l@{}}Choquette-Choo \etal{} \cite{choquette2021label}\end{tabular}                                                                        & \begin{tabular}[c]{@{}l@{}}\textit{\textbf{ML Attack Staging}} (Craft Adversarial Data),\\\textit{\textbf{Exfiltration}} (Exfiltration via ML Inference\\ API: Infer Training Data Membership)\end{tabular}                                                                                                                                                          \\ \hline
\begin{tabular}[c]{@{}l@{}}Papernot \etal{} \cite{papernot2016transferability}\end{tabular}                                                                        & \begin{tabular}[c]{@{}l@{}}\textit{\textbf{ML Attack Staging}} (Craft Adversarial Data),\\\textit{\textbf{Defense Evasion}} (Evade ML Model),\\\textit{\textbf{Impact}} (Evade ML Model)\end{tabular}                                                                                                                                                          \\ \hline
\begin{tabular}[c]{@{}l@{}}Goodfellow \etal{} \cite{goodfellow2014generative}\end{tabular}                                                                        & \begin{tabular}[c]{@{}l@{}}\textit{\textbf{ML Attack Staging}} (Craft Adversarial Data),\\\textit{\textbf{Defense Evasion}} (Evade ML Model),\\\textit{\textbf{Impact}} (Evade ML Model)\end{tabular}                                                                                                                                                         \\ \hline
\begin{tabular}[c]{@{}l@{}}Papernot \etal{} \cite{papernot2017practical}\end{tabular}                                                                        & \begin{tabular}[c]{@{}l@{}}\textit{\textbf{ML Attack Staging}} (Craft Adversarial Data),\\\textit{\textbf{Defense Evasion}} (Evade ML Model),\\\textit{\textbf{Impact}} (Evade ML Model)\end{tabular}                                                                                                                                                        \\ \hline
\begin{tabular}[c]{@{}l@{}}Cisse \etal{} \cite{cisse2017parseval}\end{tabular}                                                                        & \begin{tabular}[c]{@{}l@{}}\textit{\textbf{ML Attack Staging}} (Craft Adversarial Data)\end{tabular}                                                                                                                                                          \\ \hline
\begin{tabular}[c]{@{}l@{}}Athalye \etal{} \cite{athalye2018obfuscated}\end{tabular}                                                                        & \begin{tabular}[c]{@{}l@{}}\textit{\textbf{ML Attack Staging}} (Craft Adversarial Data),\\\textit{\textbf{Defense Evasion}} (Evade ML Model),\\\textit{\textbf{Impact}} (Evade ML Model)\end{tabular}                                                                                                                                                           \\ \hline
\begin{tabular}[c]{@{}l@{}}Jagielski \etal{} \cite{jagielski2020high}\end{tabular}                                                                        & \begin{tabular}[c]{@{}l@{}}\textit{\textbf{Reconnaissance}} (Search for Victim's \\Publicly Available Research Materials),\\\textit{\textbf{ML Attack Staging}} (Craft Adversarial Data),\\\textit{\textbf{Exfiltration}} (Exfiltration via ML Inference\\ API: Extract ML Model)\end{tabular}                                                                                                                                                           \\ \hline
\end{tabular}
}
\end{table}

\textbf{Literature}. We have manually extracted TTPs in the 14 papers by reading the attack information (i.e., technique/tactic, goal, knowledge, specificity) and linking them with TTP definitions in ATLAS and ATT\&CK. Table~\ref{new_att_lit} shows a description of the ML threats extracted and their TTPs in the form \textit{\textbf{atlas\_tactic}(atlas\_tech1, atlas\_tech2,...)}, where \textit{\textbf{atlas\_tactic}} (resp. \textit{atlas\_tech}) is an ATLAS tactic (resp. technique).

In Table~\ref{new_att_lit}, let consider the Carlini \etal{} \cite{carlini2021extracting} attack. The attack uses the GPT-2 pre-trained proxy model and then steals/extracts training examples by querying the target model. We can deduce two TTPs: ML Attack Staging (Create Proxy ML model: Use Pre-Trained Model) and Exfiltration (Exfiltration via ML Inference API: Extract ML Model). Carlini \etal{} \cite{carlini2021extracting} attack targets the privacy/confidentiality of the training data. The attack is executed in black-box mode and it tries to recover data  used to train the model. The attack also indiscriminately extracts training data (adversarially untargeted) and do not aim to extract targeted pieces of training data.

\textbf{GitHub}. We have mined 
titles and 
comments from 
issues in the 845 ML repositories and automatically extracted relevant vulnerability and threat information using two filters. The first one is the \textit{comment} filter which is a disjunction \textit{OR} of the GitHub search pattern $\langle keyword \rangle$ \textit{in:comments}, where 
$\langle keyword \rangle$ are query keywords such as \textit{cve} for CVE code, \textit{vuln} for vulnerability, \textit{threat} for threat, \textit{attack} for attack and attacker, \textit{secur} for secure and security. These signatures are commonly used by cybersecurity teams to describe potential security issues, threats, or vulnerabilities. The next one is \textit{title} filter where \textit{in:comments} is just replaced by \textit{in:title}~\cite{tech-report-v1}. 
For example, Table~\ref{cve-pattern} shows a sample of comments and titles from a filtered repository (i.e., TensorFlow) containing keyword \textit{vuln} highlighted in green color. After filtering, it remained 34 repositories  described in the repository package~\cite{tech-report-v1}. 23 repositories contained no relevant information (e.g., uncomplete CVE codes such as CVE-2018, empty contents) and have been removed. 
Finally, data from the 11 following repositories have been used for extraction: TensorFlow, OpenCV, Ray, NNI, Gym, Scikit-learn, Mxnet, Mlflow, Pytorch, Keras, Deeplearning4j~\cite{tech-report-v1}.

From the remaining data, CVE IDs are automatically extracted using the regular expression \textit{CVE-\textbackslash{}d\{4\}-\textbackslash{}d\{4,7\}}. Expression \textit{\textbackslash{}d\{4\}} matches the publication year and \textit{\textbackslash{}d\{4,7\}} matches a unique number of size between 4 and 7. Table~\ref{cve-pattern} shows a sample of comments and titles containing CVE IDs. The gray color highlights search pattern \textit{vuln} and CVE IDs found. However, the contents of titles and comments can contain threat descriptions without CVE IDs but only external links that point to other links, mailing lists, discussion threads, or other GitHub issues containing CVE IDs (e.g., Red Hat bugzilla, openwall oss-security, Apache mailing lists). Thus, we semi-automatically browsed these descriptions/sub-links/mailing lists/websites to accurately extract those hidden CVE IDs; because doing it only automatically generated a lot of false positives. 

\begin{table}[]
\caption{Tensorflow issue titles \& comments (clear CVE)}
\label{cve-pattern}
\resizebox{\columnwidth}{!}{
\begin{tabular}{|l|l|}
\hline
\textbf{titles} & \textbf{comments} \\ \hline
\begin{tabular}[c]{@{}l@{}}Upgrade Sqlite3 to \\fix \colorbox{lightgray}{CVE-2021-20227}\end{tabular} & \begin{tabular}[c]{@{}l@{}}The latest master branch of tensorflow \\ uses sqlite 3.35.5 so I think this \\\colorbox{lightgray}{vuln}erability \colorbox{lightgray}{CVE-2021-20227} ...\end{tabular} \\ \hline
\begin{tabular}[c]{@{}l@{}}segfault in tf.image.\\crop\_and\_resize \\when boxes contains \\large value\end{tabular} & \begin{tabular}[c]{@{}l@{}}...Does this issue affect the 1.15.5 version $?$\\... CVE \colorbox{lightgray}{Vuln}erability \colorbox{lightgray}{CVE-2020-15266}\\ ...patching 1.15 and 2.0 is extremely\\ expensive we no longer patched it\end{tabular} \\ \hline
\end{tabular}
}
\end{table}

\begin{table}[]
\caption{Pytorch issue titles and comments (hidden CVE)}
\label{linked-pattern}
\resizebox{\columnwidth}{!}{
\begin{tabular}{|l|l|}
\hline
\textbf{titles} & \textbf{comments} \\ \hline
\begin{tabular}[c]{@{}l@{}}pickle is a \\\colorbox{lightgray}{secur}ity issue \end{tabular}& \begin{tabular}[c]{@{}l@{}}Related: \textbf{https://github.com/pytorch/pytorch/}\\\textbf{issues/52181} ...Deprecating doesn\'t seem \\warranted; the Python pickle module itself (\\\textbf{https://docs.python.org/3/library/pickle.html})\\...Also IMHO it was a big mistake to introduce pickle\end{tabular} \\ \hline
\begin{tabular}[c]{@{}l@{}}Torch native\\\_layer\_norm OP \\out-of-bounds\\ access\end{tabular} & \begin{tabular}[c]{@{}l@{}}....Signature of native\_layer\_norm changed...\\I get a correct error message: In [3]: import torch\\...RuntimeError: Expected weight to be of same\\ shape as normalized\_shape\end{tabular} \\ \hline
\end{tabular}
}
\end{table}

\begin{figure}[h]
\centering
\includegraphics[width=0.47\textwidth]{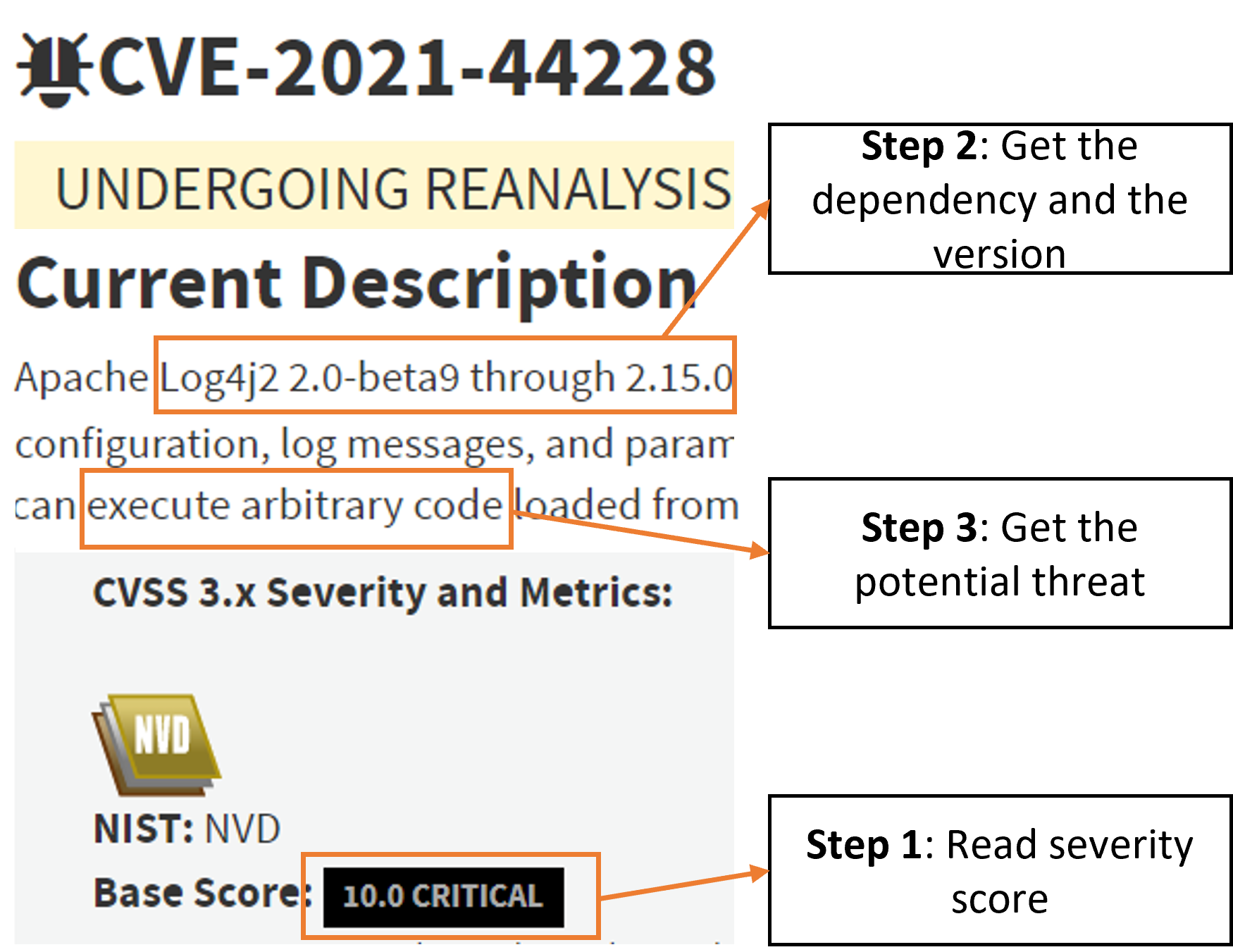}
\caption{Illustration of an investigation on the NVD database}
\label{nvd-investigation}
\end{figure}

\begin{table*}[h]
\caption{\label{tactic2phase-mapping} Mapping between tactics and ML phases}
\centering
\resizebox{\textwidth}{!}{
\begin{tabular}{|l|l|l|l|l|l|l|l|}
\hline
\backslashbox{Tactics~\cite{mitreatlas}}{ML Phases} & \textbf{Data Collection} & \textbf{\begin{tabular}[c]{@{}l@{}}Preprocessing\end{tabular}} & \textbf{\begin{tabular}[c]{@{}l@{}}Feature Engineering\end{tabular}} & \textbf{Training} & \textbf{Testing} & \textbf{Inference} & \textbf{Monitoring} \\ \hline
Reconnaissance                                                                                       & \checkmark                        & \checkmark                                                                       & \checkmark                                                                             & \checkmark                 & \checkmark                & \checkmark                  & \checkmark                \\ \hline
Resource Development                                                                                      & \checkmark                        & \checkmark                                                                       & \checkmark                                                                             & \checkmark                 & \checkmark                & \checkmark                  &                  \\ \hline
Initial Access                                                                                      & \checkmark                        &                                                                         &                                                                               & \checkmark                 & \checkmark                & \checkmark                  &                  \\ \hline
ML Model Access                                                                                   &                          &                                                                         &                                                                               & \checkmark                 & \checkmark                & \checkmark                  &                  \\ \hline
Execution                                                                                       &                          &                                                                         &                                                                               &                   & \checkmark                & \checkmark                  &                  \\ \hline
Persistence                                                                                       &                          &                                                                         &                                                                               & \checkmark                 & \checkmark                & \checkmark                  &                  \\ \hline
Defense Evasion                                                                                       & \checkmark                        & \checkmark                                                                       & \checkmark                                                                             & \checkmark                 & \checkmark                & \checkmark                  &                  \\ \hline
Discovery                                                                                       &                          &                                                                         &                                                                               & \checkmark                 & \checkmark                & \checkmark                  & \checkmark                \\ \hline
Collection                                                                                       & \checkmark                        &                                                                         &                                                                               &                   &                  &                    &                  \\ \hline
ML Attack Staging                                                                                   & \checkmark                        & \checkmark                                                                       & \checkmark                                                                             & \checkmark                 & \checkmark                & \checkmark                  &                  \\ \hline
Exfiltration                                                                                       &                          &                                                                       &                                                                               & \checkmark                 & \checkmark                & \checkmark                  &                  \\ \hline
Impact                                                                                       & \checkmark                        & \checkmark                                                                       & \checkmark                                                                             & \checkmark                 & \checkmark                & \checkmark                  & \checkmark                \\ \hline
\end{tabular}
}
\end{table*}
Table~\ref{linked-pattern} shows a sample of comments and titles from Pytorch issues that contain hidden CVE IDs. The green color highlights search pattern \textit{secur} and bold font highlights links/descriptions that allow to retrieve 
related CVE IDs. The links show that pickle affects \textit{torch.load} and \textit{torch.save} calls and mentioned a similar one \textit{numpy.load} with CVE ID: CVE-2019-6446. In total, 196 CVE IDs have been extracted from the repository issues~\cite{tech-report-v1}. We entered CVE codes into the National Vulnerability Database (NVD) and the CVE database to retrieve detailed information such as the severity level of the vulnerability (Step 1), the dependency that caused it, the version (Step 2), and the potential threats that the vulnerability can cause (Step 3). Fig.~\ref{nvd-investigation} shows an illustration of this process for the vulnerability CVE-2021-44228 caused by the Log4j dependency (version 2.15, critical score) allowing arbitrary code execution attacks.

\begin{table*}[]
\caption{\label{tactic2attack-mapping} Mapping between TTP features and attack scenarios}
\resizebox{\textwidth}{!}{
%
                                                                                     & \multicolumn{1}{l|}{stage 0}                                                                   & \multicolumn{1}{l|}{}                                                                     & \multicolumn{1}{l|}{}                                                                   & \multicolumn{1}{l|}{}                                                                   & \multicolumn{1}{l|}{}                                                              & \multicolumn{1}{l|}{}                                                                & \multicolumn{1}{l|}{}                                                                   & \multicolumn{1}{l|}{}                                                               & \multicolumn{1}{l|}{}                                                               & \multicolumn{1}{l|}{stage 1}                                                                     & \multicolumn{1}{l|}{stage 2}                                                                  &                                                            \\ \hline
\end{tabular}
}
\end{table*}

From the PyPA database, we have also extracted 33 CVE IDs in total from the 9 ML repositories using the previously defined CVE regex (i.e., \textit{CVE-\textbackslash{}d\{4\}-\textbackslash{}d\{4,7\}}). In the PyPA database, CVE codes are recorded in the YAML format PYSEC-\textit{YYYY}-\textit{UID}, where \textit{YYYY} is the publication year and \textit{UID} is a unique number. The CVE IDs that were already found in the comments/titles of the 11 remaining repositories from the GitHub search were ignored. In total, we have extracted 226 CVE IDs from the PyPA database and GitHub search. 
From the extracted data, we semi-automatically build an attack matrix for analysis as follows.

\begin{figure}[h]
\centering
\includegraphics[width=0.48\textwidth]{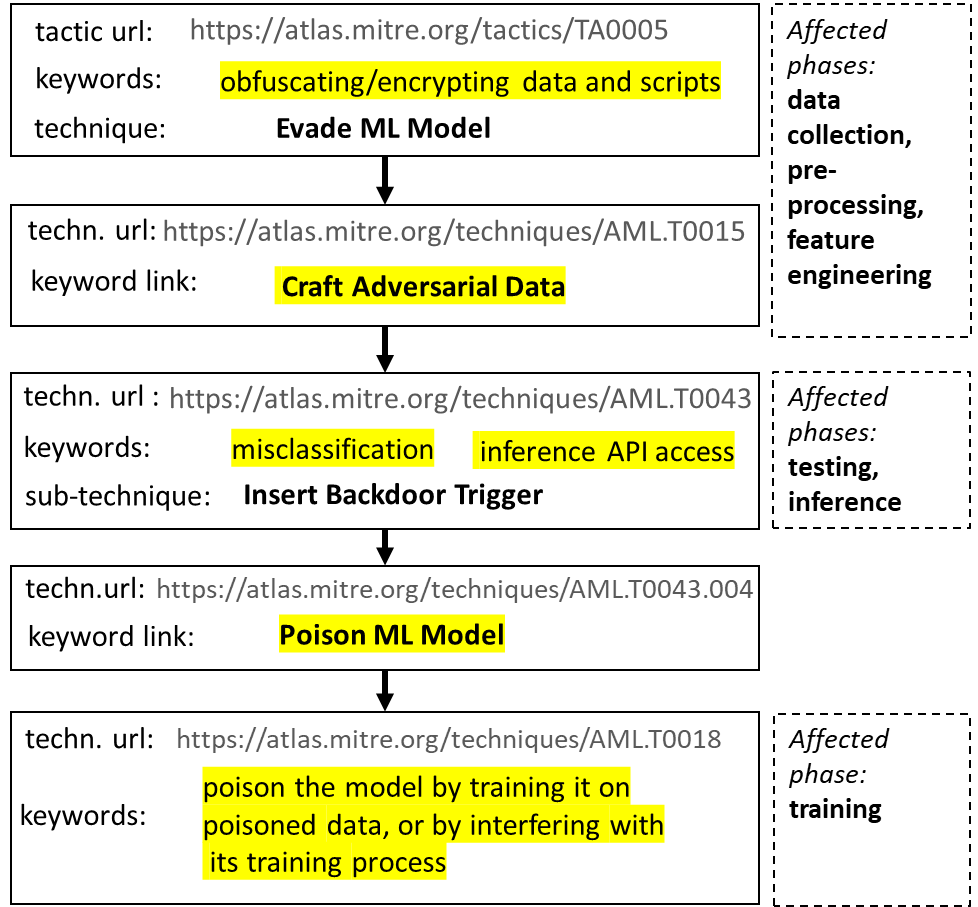}
\caption{An illustration of the process for linking tactics to ML phases}
\label{linking-mlphases}
\end{figure}

\subsubsection{Attack correlation matrix}\label{mapping-m}

After extraction, attack vectors such as vulnerability and threat are mapped to ML components (models, phases, tools). ML component is represented as a column vector while vulnerability/threat is represented as a row vector. The attack cross-correlation matrix (CCM) is the relation that maps the features of an attack vector to the features of an element-of-interest (EOI) such as ML component (phases, models, tools) or a stage of an attack scenario. For example, Table~\ref{tactic2attack-mapping} shows an attack matrix that maps TTP features (goals, knowledge, specificity, capability/tactic) to attack scenarios~\cite{mitreatlascase}. The cell \textit{('GPT-2 Model Replication', 'Reconnaissance')} is filled by $stage\;0$; it means the tactic \textit{Reconnaissance} is used by attack \textit{GPT-2 Model Replication}. The full table will be described later.  
Attack CCM is particularly important to analyze the specific 
effects that an attack vector has on a target component. Attack CCM is divided into two categories: threat CCM and vulnerability CCM. Each category is described below.\\
\textit{\underline{Threat CCM: }}
The matrix maps threat TTPs to EOIs like attack scenarios and ML phases. 

\textbf{Mapping between TTP features and attack scenarios.} In order to provide findings for RQ1, we have related threat features (i.e., goals, knowledge, specificity, capability/tactic) of ML threats to attack scenarios for identifying the most used tactics across attacks, similarities in attack execution flows and common entry points. Following the extraction process in Section~\ref{dextract}, we build the threat matrix as shown in Table~\ref{tactic2attack-mapping}. The coefficients of this matrix can take a string value (e.g., Black-box, stage 0), an empty value when there is no relation between the feature and the attack scenario, or an \textsf{N/A} value meaning that the relation between the feature and the attack scenario is unknown or not mentioned in the database. In the attack capability column, $stage\;i$ means that the attack scenario executes a given tactic at step $i$ in the attack execution flow. $stage\;i,stage\;j$ means that $stage\;i$ and $stage\;j$ executes the same tactic. 

For example, attack \textit{VirusTotal Poisoning} starts its execution at stage $stage\;0$ 
(i.e., Resource Development) in order to purchase tools or infrastructure to support operations. Next, it executes stage $stage\;1$ 
(i.e., ML Attack Staging) for crafting adversarial data and poison the target model. Then, attacker executes stage $stage\;2$ 
(i.e., Initial Access) for exploiting valid accounts or external remote services to gain access to unauthorized resources. The attack scenario stops at stage $stage\;3$ (i.e., Persistence) to keep access on the target. 


\textbf{Mapping between tactics and ML phases.} In order to answer RQ2, tactics are also mapped to ML phases for identifying the frequent threat tactics used against each ML phase. Firstly, we read the ATLAS description of each tactic and their related techniques to find ML phase signatures such as \textit{trained}, \textit{testing the model}, \textit{use statistics of model prediction scores}, and \textit{obfuscating/encrypting data}. Next, we associate keywords (e.g., \textit{inference API}) to ML phases (e.g., Inference) following the attack description. An illustration of this process for tactic \textit{Defense Evasion} is shown in Fig.~\ref{linking-mlphases}. The tactic URL shows a  keyword \textit{obfuscating/encrypting data and scripts} that clearly affects data manipulation phases such as data collection, pre-processing, and feature engineering. When going further by opening technique \textit{Craft Adversarial Data}, the technique URL shows that attack can cause \textit{misclassification} with adversarial examples and get \textit{inference API access}, which clearly affects testing and inference phases. By going deeper with technique \textit{Poison ML Model}, it is clearly shown that the attack affects the training phase. Then, the relationship between tactics and ML phases are recorded in a threat CCM for analysis.
Table \ref{tactic2phase-mapping} shows a record of the mapping between tactics and ML phases to figure out the impact of threat TTPs against ML phases. The coefficients of this matrix are represented by symbol $\checkmark$ when there is a relation between a given tactic and ML phase. Results for RQ2 are provided in Section~\ref{evaluation}.

\begin{table}[]
\caption{\label{attack2model-mapping} Mapping between attack scenarios and ML models}
\resizebox{\columnwidth}{!}{
\begin{tabular}{|l|l|l|}
\hline
\multirow{2}{*}{\textbf{Source}}                                                            & \multirow{2}{*}{\textbf{\begin{tabular}[c]{@{}l@{}}Attack\\ scenario\end{tabular}}}                                                            & \multirow{2}{*}{\textbf{\begin{tabular}[c]{@{}l@{}}Model\\ Used\end{tabular}}} \\
                                                                                            &                                                                                                                                                &                                                                                \\ \hline
\multirow{15}{*}{\textbf{\begin{tabular}[c]{@{}l@{}}MITRE \\ \\ ATLAS\end{tabular}}}        & \begin{tabular}[c]{@{}l@{}}Evasion of Deep Learning\\ Detector for Malware C2\\ Traffic\end{tabular}                                           & CNN                                                                            \\ \cline{2-3} 
                                                                                            & \begin{tabular}[c]{@{}l@{}}Botnet Domain Generation\\ (DGA)  Detection Evasion\end{tabular}                                                    & CNN                                                                            \\ \cline{2-3} 
                                                                                            & VirusTotal Poisoning                                                                                                                           & LSTM                                                                           \\ \cline{2-3} 
                                                                                            & \begin{tabular}[c]{@{}l@{}}Bypassing Cylance's AI \\ Malware Detection\end{tabular}                                                            & \begin{tabular}[c]{@{}l@{}}DNN\end{tabular}                    \\ \cline{2-3} 
                                                                                            & \begin{tabular}[c]{@{}l@{}}Camera Hijack Attack on \\ Facial Recognition System\end{tabular}                                                   & CNN, GAN                                                                       \\ \cline{2-3} 
                                                                                            & \begin{tabular}[c]{@{}l@{}}Attack on Machine\\ Translation  Service - \\Google Translate, Bing \\ Translator, and \\Systran Translate\end{tabular} & \begin{tabular}[c]{@{}l@{}}Transformer\end{tabular}            \\ \cline{2-3} 
                                                                                            & \begin{tabular}[c]{@{}l@{}}Clearview AI \\Misconfiguration\end{tabular}                                                                                                                  & N/A                                                                            \\ \cline{2-3} 
                                                                                            & GPT-2 Model Replication                                                                                                                        & GPT-2                                                                          \\ \cline{2-3} 
                                                                                            & ProofPoint Evasion                                                                                                                             & Copycat~\cite{correia2018copycat}                                                                            \\ \cline{2-3} 
                                                                                            & Tay Poisoning                                                                                                                                  & \begin{tabular}[c]{@{}l@{}}DNN\end{tabular}                   \\ \cline{2-3} 
                                                                                            & \begin{tabular}[c]{@{}l@{}}Microsoft Azure \\Service Disruption\end{tabular}                                                                                                             & N/A                                                                            \\ \cline{2-3} 
                                                                                            & Microsoft Edge AI Evasion                                                                                                                      & \begin{tabular}[c]{@{}l@{}}DNN\end{tabular}                    \\ \cline{2-3} 
                                                                                            & \begin{tabular}[c]{@{}l@{}}Face Identification System \\ Evasion via Physical Counter\\ measures\end{tabular}                                  & N/A                                                                            \\ \cline{2-3} 
                                                                                            & \begin{tabular}[c]{@{}l@{}}Backdoor Attack on\\ Deep Learning Models\\ in Mobile Apps\end{tabular}                                              & \begin{tabular}[c]{@{}l@{}}DNN\end{tabular}                    \\ \cline{2-3} 
                                                                                            & \begin{tabular}[c]{@{}l@{}}Confusing AntiMalware \\ Neural Networks\end{tabular}                                                               & \begin{tabular}[c]{@{}l@{}}DNN\end{tabular}                    \\ \hline
\multirow{8}{*}{\textbf{\begin{tabular}[c]{@{}l@{}}AI \\ Incident\\ Database\end{tabular}}} & \begin{tabular}[c]{@{}l@{}}India's Tek Fog Shrouds an \\ Escalating Political War\end{tabular}                                                 & GPT-2                                                                          \\ \cline{2-3} 
                                                                                            & \begin{tabular}[c]{@{}l@{}}Meta Says It's Shut Down A\\ Pro-Russian DisInformation\\ Network...\end{tabular}                                    & N/A                                                                            \\ \cline{2-3} 
                                                                                            & \begin{tabular}[c]{@{}l@{}}Libyan Fighters Attacked by a\\ Potentially Unaided Drone, \\ UN Says\end{tabular}                                  & CNN                                                                            \\ \cline{2-3} 
                                                                                            & \begin{tabular}[c]{@{}l@{}}Fraudsters Cloned Company\\ Director's Voice In \$35M Bank \\ Heist, Police Find\end{tabular}                       & DeepVoice~\cite{arik2017deep}                                                                      \\ \cline{2-3} 
                                                                                            & \begin{tabular}[c]{@{}l@{}}Poachers Evade KZN Park's \\ High-Tech Security\\ and Kill four \\ Rhinos for their Horns\end{tabular}                & \begin{tabular}[c]{@{}l@{}}DNN\end{tabular}                    \\ \cline{2-3} 
                                                                                            & \begin{tabular}[c]{@{}l@{}}Tencent Keen Security Lab:\\ Experimental Security \\Research of Tesla Autopilot\end{tabular}                      & Fisheye~\cite{tencent-security-lab}                                                                            \\ \cline{2-3} 
                                                                                            & \begin{tabular}[c]{@{}l@{}}Three Small Stickers\\ in Intersection Can\\ Cause Tesla Autopilot to \\ Swerve Into Wrong Lane\end{tabular}         & CNN                                                                            \\ \cline{2-3} 
                                                                                            & \begin{tabular}[c]{@{}l@{}}The DAO Hack -Stolen \$50M \\ The Hard Fork\end{tabular}                                                            & N/A                                                                            \\ \hline
\multirow{22}{*}{\textbf{Literature}}                                                       & Carlini \etal{}~\cite{carlini2021extracting}                                                                                                                                 & GPT-2                                                                          \\ \cline{2-3} 
                                                                                            & Biggio \etal{} \cite{biggio2013evasion}                                                                                                                                & SVM, DNN                                                                               \\ \cline{2-3} 
                                                                                            & Barreno \etal{}~\cite{barreno2010security}                                                                                                                               &  Naive Bayes                                                                              \\ \cline{2-3} 
                                                                                            & Carlini \etal{}~\cite{carlini2017adversarial}                                                                                                                                   &   Feed-Forward DNN                                                                             \\ \cline{2-3} 
                                                                                            & Wallace \etal{} \cite{wallace2020imitation}                                                                                                                               & Transformer                                                                                \\ \cline{2-3} 
                                                                                            & Abdullah \etal{} \cite{abdullah2021sok}                                                                                                                                  &  \begin{tabular}[c]{@{}l@{}}RNN, CNN,\\Hidden Markov\end{tabular}                                                                              \\ \cline{2-3} 
                                                                                            & Chen \etal{} \cite{chen2021badnl}                                                                                                                                  &  LSTM, BERT      \\ \cline{2-3} 
                                                                                            & Choquette-Choo \etal{} \cite{choquette2021label}                                                                                                                                  &  CNN, RestNet                                                                              \\ \cline{2-3} 
                                                                                            & Papernot \etal{}~\cite{papernot2016transferability}                                                                                                                                 & \begin{tabular}[c]{@{}l@{}}DNN, kNN, SVM,\\ Logistic Regression\\ Decision Trees\end{tabular}                 \\ \cline{2-3} 
                                                                                            & Goodfellow \etal{} \cite{goodfellow2014generative}                                                                                                                                 &  GAN                                                                              \\ \cline{2-3} 
                                                                                            &  Papernot \etal{}~\cite{papernot2017practical}                                                                                                                                &  DNN                                                                         \\ \cline{2-3} 
                                                                                            & Cisse \etal{}~\cite{cisse2017parseval}                                                                                                                          &    Parseval Networks                                                                            \\ \cline{2-3} 
                                                                                            & Athalye \etal{}~\cite{athalye2018obfuscated}                                                                                                                                   &  \begin{tabular}[c]{@{}l@{}}CNN, ResNet,\\ InceptionV3\end{tabular}                                                                               \\ \cline{2-3} 
                                                                                            & Jagielski \etal{} \cite{jagielski2020high}                                                                                                                                 & RestNetv2                 \\ \hline
\end{tabular}
}
\end{table}

\textbf{Mapping between attack scenarios and ML models.} To identify ML models targeted/exploited by attack scenarios (as stated in RQ2), the mapping process is done by searching the model type exploited for the attack or a similar attack in the ATLAS TTP descriptions and the attack descriptions from the AI Incident database. When nothing is found, we google the name of the attack scenario and then read related public news and research papers to get the model type of the attack. For instance, the description of the attack scenario \textit{Botnet DGA Detection Evasion} 
\begin{displayquote}
The Palo Alto Networks Security AI research team was able to bypass a \textbf{Convolutional Neural Network (CNN)}-based botnet Domain Generation Algorithm (DGA) detection [1] by domain name mutations.
\end{displayquote}
indicates that the target model is Convolutional Neural Network~\cite{mitreatlascase}. Next, the description of the attack scenario \textit{Attack on Machine Translation Service} 
\begin{displayquote}
...A research group at UC Berkeley utilized these public endpoints to create an replicated model with near-production...These adversarial inputs ...dropped sentences on Google Translate and Systran Translate websites.
\end{displayquote}
do not contain information about the replication model used~\cite{mitreatlascase}. Thus, we google the attack scenario title and it pointed to the paper~\cite{wallace2020imitation}. This paper allowed us to figure out that the replication was based on the Transformer model and Knowledge Distillation~\cite{wallace2020imitation}. Then, the targeted/exploited ML models and attack scenarios are recorded in a threat CCM for analysis. 

\begin{table*}[]
\centering
\caption{An illustration of the generated vulnerability CCM}
\label{vuln-ccm}
\resizebox{\textwidth}{!}{
\begin{tabular}{|l|l|l|l|l|l|l|l|l|l|l|l|}
\hline
\textbf{\backslashbox{\textbf{CVE ID}}{\textbf{ML tool}}} & \textbf{tensorflow}                                                                 & \textbf{opencv} & \textbf{ray}                                                                              & \textbf{nni}                                                                             & \textbf{gym}        & \textbf{scikit-learn}                                                                    & \textbf{mxnet}                                                                              & \textbf{mlflow}         & \textbf{pytorch}                                                                  & \textbf{keras}                                                                         & \textbf{deeplearning4j}                                                          \\ \hline
\textbf{CVE-2022-29216}             & \begin{tabular}[c]{@{}l@{}}(this,code \\injection,high)\end{tabular}                                                        &                 &                                                                                           &                                                                                          &                     &                                                                                          &                                                                                             &                         &                                                                                   &                                                                                        &                                                                                  \\ \hline
\textbf{CVE-2022-29213}             & \begin{tabular}[c]{@{}l@{}}(this,improper input\\ validation, medium)\end{tabular} &                 &                                                                                           &                                                                                          &                     &                                                                                          &                                                                                             &                         &                                                                                   &                                                                                        &                                                                                  \\ \hline
\textbf{CVE-2022-29208}             & \begin{tabular}[c]{@{}l@{}}(this,out-of-bounds\\write,high)\end{tabular}                                                       &                 &                                                                                           &                                                                                          &                     &                                                                                          &                                                                                             &                         &                                                                                   &                                                                                        &                                                                                  \\ \hline
\textbf{CVE-2022-29203}             & \begin{tabular}[c]{@{}l@{}}(this,integer \\overflow,medium) \end{tabular}                                                       &                 &                                                                                           &                                                                                          &                     &                                                                                          &                                                                                             &                         &                                                                                   &                                                                                        &                                                                                  \\ \hline
\textbf{CVE-2022-21741}             & \begin{tabular}[c]{@{}l@{}}(this,divide \\by zero,medium)\end{tabular}                                                        &                 &                                                                                           &                                                                                          &                     &                                                                                          &                                                                                             &                         &                                                                                   &                                                                                        &                                                                                  \\ \hline
\textbf{CVE-2021-37650}             & \begin{tabular}[c]{@{}l@{}}(this,out-of\\-bounds write,high) \end{tabular}                                                    &                 &                                                                                           &                                                                                          &                     &                                                                                          &                                                                                             &                         &                                                                                   &                                                                                        &                                                                                  \\ \hline
\textbf{CVE-2021-37648}             & \begin{tabular}[c]{@{}l@{}}(this,null pointer \\ dereference,high)\end{tabular}     &                 &                                                                                           &                                                                                          &                     &                                                                                          &                                                                                             &                         &                                                                                   &                                                                                        &                                                                                  \\ \hline
\textbf{CVE-2021-44832}             &                                                                                     &                 & (log4j, rce,  medium)                                                                     &                                                                                          &                     &                                                                                          &                                                                                             &                         &                                                                                   &                                                                                        &                                                                                  \\ \hline
\textbf{CVE-2021-44228}             &                                                                                     &                 & (log4j, rce,  critical)                                                                   &                                                                                          &                     &                                                                                          &                                                                                             & (log4j, rce,  critical) &                                                                                   &                                                                                        &                                                                                  \\ \hline
\textbf{CVE-2021-3177}              &                                                                                     &                 &                                                                                           &                                                                                          &                     &                                                                                          &                                                                                             &                         & \begin{tabular}[c]{@{}l@{}}(python391, buffer\\ -overflow, critical)\end{tabular} &                                                                                        &                                                                                  \\ \hline
\textbf{CVE-2021-27921}             &                                                                                     &                 &                                                                                           &                                                                                          & (pillow, dos, high) &                                                                                          &                                                                                             &                         &                                                                                   &                                                                                        &                                                                                  \\ \hline
\textbf{CVE-2020-28975}             &                                                                                     &                 &                                                                                           &                                                                                          &                     & (libsvm, dos, high)                                                                      &                                                                                             &                         &                                                                                   &                                                                                        &                                                                                  \\ \hline
\textbf{CVE-2020-24342}             &                                                                                     &                 & \begin{tabular}[c]{@{}l@{}}(lua54,  memo-buffer-\\ restriction-error,  high)\end{tabular} &                                                                                          &                     &                                                                                          &                                                                                             &                         &                                                                                   &                                                                                        &                                                                                  \\ \hline
\textbf{CVE-2020-15945}             &                                                                                     &                 & \begin{tabular}[c]{@{}l@{}}(lua54,  out-of-\\bound-memory-\\access, medium)\end{tabular}   &                                                                                          &                     &                                                                                          &                                                                                             &                         &                                                                                   &                                                                                        &                                                                                  \\ \hline
\textbf{CVE-2020-13092}             &                                                                                     &                 &                                                                                           &                                                                                          &                     & \begin{tabular}[c]{@{}l@{}}(joblib, untrusted-de\\serialization, critical)\end{tabular} &                                                                                             &                         &                                                                                   &                                                                                        &                                                                                  \\ \hline
\textbf{CVE-2020-15208}             & \begin{tabular}[c]{@{}l@{}}(this,out-of-bounds \\ read,critical)\end{tabular}       &                 &                                                                                           &                                                                                          &                     &                                                                                          &                                                                                             &                         &                                                                                   &                                                                                        &                                                                                  \\ \hline
\textbf{CVE-2020-15206}             & \begin{tabular}[c]{@{}l@{}}(this,improper input\\ validation,high)\end{tabular}    &                 &                                                                                           &                                                                                          &                     &                                                                                          &                                                                                             &                         &                                                                                   &                                                                                        &                                                                                  \\ \hline
\textbf{CVE-2020-25649}             &                                                                                     &                 &                                                                                           &                                                                                          &                     &                                                                                          &                                                                                             &                         &                                                                                   &                                                                                        & \begin{tabular}[c]{@{}l@{}}(jackson-bind100,\\ xxe-injection, high)\end{tabular} \\ \hline
\textbf{CVE-2020-11022}             &                                                                                     &                 &                                                                                           &                                                                                          &                     &                                                                                          &                                                                                             &                         &                                                                                   & \begin{tabular}[c]{@{}l@{}}(jquery211, cross-site \\ -scripting,  medium)\end{tabular} &                                                                                  \\ \hline
\textbf{CVE-2019-6446}              &                                                                                     &                 &                                                                                           & \begin{tabular}[c]{@{}l@{}}(pickle,  arbritary\\-code-execution,\\ critical)\end{tabular} &                     &                                                                                          &                                                                                             &                         &                                                                                   &                                                                                        &                                                                                  \\ \hline
\textbf{CVE-2018-18074}             &                                                                                     &                 &                                                                                           &                                                                                          &                     &                                                                                          & \begin{tabular}[c]{@{}l@{}}(requests,\\credentials-snif\\fing, high)\end{tabular} &                         &                                                                                   &                                                                                        &                                                                                  \\ \hline
\end{tabular}
}
\end{table*}

Table \ref{attack2model-mapping} shows the mapping between attack scenarios and ML models per dataset to figure out the impact of threat TTPs against ML models. Attack scenarios exploited/targeted 16 specific model architectures such as Convolutional Neural Network (CNN), Long Short Term Memories (LSTM), Generative Adversarial Network (GAN), Generative Pretrained Transformer (GPT), DeepVoice~\cite{arik2017deep}, Support Vector Machine (SVM), Recurrent Neural Network (RNN), Feed-Forward Neural Network (FFNN), Copycat CNN~\cite{correia2018copycat}, Residual Neural Network (ResNet), Fisheye CNN~\cite{tencent-security-lab}, Logistic Regression, k-Nearest Neighbor (kNN), Decision Tree (DT), Parseval Network, InceptionV3, Hidden Markov Model (HMM), Naive Bayes, and Bidirectional Encoder Representations from Transformer (BERT). 
Other ML attacks exploited/targeted Deep Neural Networks (DNNs) or Transformers but the type is unknown. Some model architectures used by threat actors were unknown after several researches and the notation \textit{N/A} is used in such case. Results for RQ2 are provided in Section~\ref{evaluation}.\\

\textit{\underline{Vulnerability CCM}}: 
The matrix maps vulnerabilities to EOIs like ML repositories.
To get findings for RQ3, CVE IDs found in repository issues and those from the PyPA database have been mapped to ML repositories for identifying the most prominent vulnerabilities and threats in ML repositories as well as the dependencies that cause them (see Fig.~\ref{nvd-investigation}). The mapping relation between an CVE ID and a given ML tool is represented by a tuple $\langle dep, att, lvl \rangle$, where $dep$ is the name of the dependency that caused the attack; $dep=this$ if it is the main tool, $att$ is the name of the attack that can be launched to exploit a CVE vulnerability, and $lvl$ is the severity level of the vulnerability. 

In Table~\ref{vuln-ccm}, an illustration of the mapping is described using vulnerability extraction process in Table~\ref{cve-pattern} and Table~\ref{linked-pattern}. The cell \textbf{(CVE-2022-29216, tensorflow)} has value \textit{(this, code injection, high)}. It means that Tensorflow framework is vulnerable to code injection with a high severity. In addition, the cell \textbf{(CVE-2019-6446, nni)} has value \textit{(pickle, arbitrary-code-exec, critical)}, meaning that the dependency \textit{pickle} has the highest severity level (i.e., critical) and it can allow arbitrary code execution (arbitrary-code-exec) attacks in the Microsoft Neural Network Intelligence (NNI) tool. The full tables can be found in \cite{tech-report-v1}.

\subsubsection{Metrics}
We compute the following metrics to answer our research questions.\\
\textbf{RQ1.} For this question, the metric used is the number of attacks that used a given tactic and the tactic. It provides information about the most exploited tactics by attack scenarios.\\
\textbf{RQ2.} The metric used is the number of tactics that target a given phase and the phase targeted. It gives information about ML phases that are more impacted/targeted by ML threats. \\
\textbf{RQ3.} For this question, the metrics are the number of vulnerabilities (nov) and the vulnerability type, the nov per tool, and the nov per type per tool. They provide information about the prominence of vulnerabilities in the ML repositories, the most affected repositories, and the potential threats.

\section{Study results}\label{evaluation}

In this section, we present and discuss the results of our research questions. 

\subsection{
 Prominence and common entry points of threat TTPs exploited in ML attack scenarios (RQ1)}
 
This research question is divided into two parts: the prominence of threat TTPs exploited in attack scenarios and the common entry points.


\subsubsection*{The prominence of threat TTPs in attack scenarios}

Table~\ref{tactic2attack-mapping} shows the mapping between tactics and attack scenarios. The most prominent tactic is \textit{ML Attack Staging}; since it occurs 30 times across the 89 ML attack scenarios. During ML attack staging, threat actors prepare their attack by crafting adversarial data to feed the target model, training proxy models, poisoning or evading the target model. The other significant tactics used in attack scenarios are \textit{Impact} and \textit{Resource Development}; since they respectively occur 21 times and 15 times in ML attack scenarios (see Table~\ref{tactic2attack-mapping}). After the ML attack success, most attack scenarios tried to evade ML model, disrupt ML service, or destroy ML systems, data, and humans (Impact). 


In Table~\ref{tactic2attack-mapping}, the execution flows of attack scenarios share some TTP stages. The most used TTP sequences in attack scenarios are:
\begin{itemize}
    \item \textit{ML Attack Staging (stage 0)} $\rightarrow$ \textit{ Defense Evasion (stage 1)} $\rightarrow$ \textit{ Impact (stage 2)}
    \item \textit{ML Attack Staging (stage 0)} $\rightarrow$ \textit{ Exfiltration (stage 1)} 
    \item \textit{ML Attack Staging (stage 0)} $\rightarrow$ \textit{ Impact (stage 1)} 
    \item \textit{Reconnaissance (stage 0)} $\rightarrow$ \textit{Resource Development (stage 1)} $\rightarrow$ \textit{ML Model Access (stage 2)} $\rightarrow$ \textit{ML Attack Staging (stage 3)} $\rightarrow$ \textit{Impact (stage 4)}
    \item \textit{Reconnaissance (stage 0)} $\rightarrow$ \textit{Resource Development (stage 1)} $\rightarrow$ \textit{ML Attack Staging (stage 2)} $\rightarrow$ \textit{Defense Evasion (stage 3)}
\end{itemize}

since attack scenarios (Carlini \etal{} \cite{carlini2017adversarial}, Abdullah \etal{} \cite{abdullah2021sok}, Papernot \etal{} \cite{papernot2016transferability, papernot2017practical}, Biggio \etal{} \cite{biggio2013evasion}, Athalye \etal{} \cite{athalye2018obfuscated}, Barreno \etal{} \cite{barreno2010security}) have similar execution sequences i.e., starting from stage \textit{stage 0} to stage \textit{stage 2}. Attack scenarios (Carlini \etal{} \cite{carlini2021extracting}, Wallace \etal{} \cite{wallace2020imitation}, Choquette-Choo \etal{} \cite{choquette2021label}) share stages from \textit{stage 0} to \textit{stage 1}. In addition, attack scenarios \textit{Attack on Machine Translation Service} and \textit{Microsoft Edge
AI Evasion} have similar execution sequences i.e., starting from stage \textit{S0} to stage \textit{S4}. It is also the same for attack scenarios \textit{Evasion of Deep Learning Detector for Malware C2 Traffic} and \textit{Botnet Domain Generation (DGA) Detection Evasion} that share stages from \textit{stage 0} to \textit{stage 3}. Attack scenarios \textit{Jagielski} \etal{} \cite{jagielski2020high} and \textit{Poachers Evade KZN's Park High-Tech Security} have some stages already included in the selected sequences, i.e., \textit{Defense Evasion} (stage 0) and \textit{Impact} (stage 1), \textit{ML Attack Staging} (stage 1) and \textit{Exfiltration} (stage 2). Attack scenarios \textit{Backdoor Attack on Deep Learning Models in Mobile Apps} and \textit{Confusing AntiMalware Neural Networks} only share two stages (i.e., \textit{stage 0} and \textit{stage 1}) already included in the selected sequences; thus, they are ignored.  

Table~\ref{tactic2attack-mapping} also shows that the most attack scenarios targeted ML systems without a prior knowledge or access of the training data and the ML model (black box); this is explained by the highest number of occurrences of \textit{Black-box} in the Attack Knowledge column (i.e., 17 times). In addition, most attack scenarios are untargeted, shown by the highest number of occurrences of \textit{Traditional Untargeted} and \textit{Adversarially Untargeted} in the Attack Specificity column (i.e., 20 times). They also mainly targeted \textit{Confidentiality} and \textit{Integrity}.

\begin{boxblock}{Summary 1}
    \begin{itemize}
      \item The common entry points of ML attacks are \textit{Reconnaissance} and \textit{ML Attack Staging}
      \item The most used TTPs by ML attacks are \textit{ML Attack Staging} and \textit{Impact}
      \item The shortest used TTP sequences are \textit{ML Attack Staging} $\rightarrow$ \textit{Exfiltration} and \textit{ML Attack Staging} $\rightarrow$ \textit{Impact} 
      \item The longest used TTP sequence is \textit{Reconnaissance} $\rightarrow$ \textit{Resource Development} $\rightarrow$ \textit{ML Model Access} $\rightarrow$ \textit{ML Attack Staging} $\rightarrow$ \textit{Impact}
      \item Most ML attack scenarios targeted \textit{Confidentiality} and \textit{Integrity}
    \end{itemize}
\end{boxblock}


\subsubsection*{The common entry points in attack scenarios} Table~\ref{tactic2attack-mapping} shows that
the common entry points of attack scenarios are \textit{Reconnaissance} and \textit{ML Attack Staging}. Precisely, attackers exploited public resources such as research materials (e.g., research papers, pre-print repositories), ML artifacts like existing pre-trained models and tools (e.g., GPT-2), and adversarial ML attack implementations (Reconnaissance). To start the attack, they can use pre-trained proxy model or craft adversarial data offline to be send to the ML model for attack (ML Attack Staging). 


\subsection{Impact of threat TTPs against ML phases and models (RQ2)}

This research question aims to identify the 
most targeted/vulnerable ML phases and models and the most used threat TTPs across different ML phases. Therefore, the question is divided into two parts: the impact of threat TTPs against ML phases and the impact of threat TTPs against ML models. 

\subsubsection*{Impact of threat TTPs against ML phases}
In Table~\ref{tactic2phase-mapping}, the most targeted phases are \textit{Testing}, \textit{Inference}, \textit{Training}, and \textit{Data collection}. Tactics \textit{Reconnaissance}, \textit{Impact}, \textit{ML Attack Staging}, and \textit{Resource Development} are the most used across the different ML phases.

\begin{table}[h]
\centering
\caption{Statistics about models used in attack scenarios}
\label{modelstat}
\begin{tabular}{|l|l|l|}
\hline
\textbf{Models used}                                                   & \textbf{\begin{tabular}[c]{@{}l@{}}Occurrences\\ per attack\end{tabular}} & \textbf{\begin{tabular}[c]{@{}l@{}}Attack\\ Period\end{tabular}} \\ \hline
\begin{tabular}[c]{@{}l@{}}Transformers \\ (BERT, GTP-2, GPT-3, others)\end{tabular} & 6                                                                         & 2019-2022                                                        \\ \hline
\begin{tabular}[c]{@{}l@{}}Convolutional Neural Networks\\ (CopyCat, Fisheye, ResNet, others)\end{tabular}                                                                    & 11                                                                         & 2018-2021                                                        \\ \hline
\begin{tabular}[c]{@{}l@{}}Deep Neural Networks\\ (unspecified)
\end{tabular}                                                      & 9                                                                         & 2013-2021                                                        \\ \hline
Hidden Markov & 1                                                                         & 2021                                                       \\ \hline
Long-Short Term Memory                                                                   & 2                                                                         & 2020-2021                                                             \\ \hline
\begin{tabular}[c]{@{}l@{}}Generative Adversarial Networks\end{tabular}                                                                    & 2                                                                         & 2014-2020                                                             \\ \hline
DeepVoice~\cite{arik2017deep}                                                             & 1                                                                         & 2019                                                             \\ \hline
Feed-Forward Neural Networks                                                          & 1                                                                         & 2017                                                             \\ \hline
\begin{tabular}[c]{@{}l@{}}Parseval Networks\end{tabular} & 1                                                                         & 2017                                                       \\ \hline
\begin{tabular}[c]{@{}l@{}}Linear classifiers \\ (SVM, Logistic Reg., Naive Bayes)\end{tabular} & 3                                                                         & 2010-2016                                                        \\ \hline
\begin{tabular}[c]{@{}l@{}}Non-Linear classifiers \\ (Decision Trees, k-Nearest Neighbor)\end{tabular} & 2                                                                         & 2016                                                        \\ \hline
N/A                                                                    & 5                                                                         & 2018-2022                                                        \\ \hline
\end{tabular}
\end{table}

\subsubsection*{Impact of threat TTPs against ML models}
Table~\ref{modelstat} shows the number of occurrences of the models used in attack scenarios and the period of the attack based on models extracted in Section~\ref{mapping-m}. \textsf{N/A} means that information about the model used is not found. The mention "unspecified" in row 2 means attack scenarios used Deep Neural Networks (DNNs) but the architecture was not specified. Among the attack scenarios, the most targeted models between 2018 and 2021 are Convolutional Neural Networks (CNNs). CNNs was targeted by 11 attack scenarios from ATLAS, AI Incident databases, and the literature including \textit{Evasion of Deep Learning detector for malware C\&C traffic}, \textit{Botnet Domain Generation Algorithm (DGA) Detection Evasion}, \textit{Camera Hijack Attack on Facial Recognition System}, \textit{Libyan Fighters Attacked by a Potentially Unaided Drone}, \textit{Tencent Keen Security Lab: Experimental Security Research of Tesla Autopilot}, \textit{Athalye} \etal{} \cite{athalye2018obfuscated}, and \textit{Abdullah} \etal{} \cite{abdullah2021sok}. Attacks such as \textit{Attack on Machine Translation Service}, \textit{GPT-2 Model Replication}, \textit{India Tek Fog Shrouds an Escalating Political War}, and \textit{Wallace} \etal{} \cite{wallace2020imitation} also targeted Transformer models~\cite{vaswani2017attention}.

In the Kaggle report 2021\mbox{~\cite{kagglereport}}, Gradient Boosting Machines (xgboost, lightgbm) are the most popular models used by the AI community followed by CNNs between 2019 and 2021. These statistics confirm the choice of CNN models by threat actors in the studied attack scenarios. However, the studied attack scenarios did not use tree-based models such as gradient boosting machines. A reason is that of algorithms for crafting adversarial examples
for tree-based models are limited due their discrete and non-differentiable nature; that restricts the execution of gradient-based white-box attacks~\cite{chen2019robust}.


\begin{boxblock}{Summary 2}
    \begin{itemize}
      \item The most used TTPs against ML phases are \textit{Reconnaissance}, \textit{Impact}, \textit{ML Attack Staging}, and \textit{Resource Development}
      \item The most targeted ML phases are \textit{Testing}, \textit{Inference}, \textit{Training}, and \textit{Data collection}
      \item CNNs are the most targeted among the attack scenarios from ATLAS, the AI Incident database, and the literature
    \end{itemize}
\end{boxblock}

\subsection{Finding new threats in the AI Incident database, the literature, and ML repositories that are not documented in ATLAS (RQ3)}


To achieve this goal, the question is splitted into three parts:
the new threats 
from the AI Incident database and the literature, and the potential threats from the ML repositories, the most vulnerable ML repositories as well as the dependencies that cause them, and the most frequent vulnerabilities 
in the ML repositories.

\subsubsection*{The new threats from the AI Incident Database and the literature}

In Table~\ref{new_att}, we have identified new TTPs (i.e., 9 techniques, 7 tactics) from the 8 ML attacks in the AI Incident database. The attack techniques used are \textit{Establish Accounts, Valid Accounts, Create Proxy ML model: Use Pre-Trained Model, Exfiltration via Cyber Means, Active Scanning, Cost Harvesting, Evade ML Model, Craft Adversarial Data,} and \textit{ML Denial of Service}. The attack tactics used are \textit{Resource Development, Initial Access, ML Attack Staging, Exfiltration, Reconnaissance, Impact,} and \textit{Defense Evasion}.

Table~\ref{new_att_lit} also shows TTPs extracted from the 14 ML attacks in literature (i.e., 8 techniques, 6 tactics) following ATLAS TTP definitions. The attack techniques used are \textit{Create Proxy ML model: Use Pre-Trained Model, Exfiltration via ML Inference
API: Extract ML Model. Craft Adversarial Data, Evade ML Model, Craft Adversarial Data:
Insert Backdoor Trigger, Backdoor ML Model:
Inject Payload, Exfiltration via ML Inference
API: Infer Training Data Membership,} and \textit{Search for Victim’s Publicly Available Research Materials}. The attack tactics used are \textit{ML Attack Staging, Exfiltration, Defense Evasion, Impact, Persistence,} and \textit{Reconnaissance}.

These 32 new ML attack scenarios were not documented in ATLAS and can be used to complete ATLAS case studies.

\begin{figure}[h]
\centering
\includegraphics[width=0.49\textwidth]{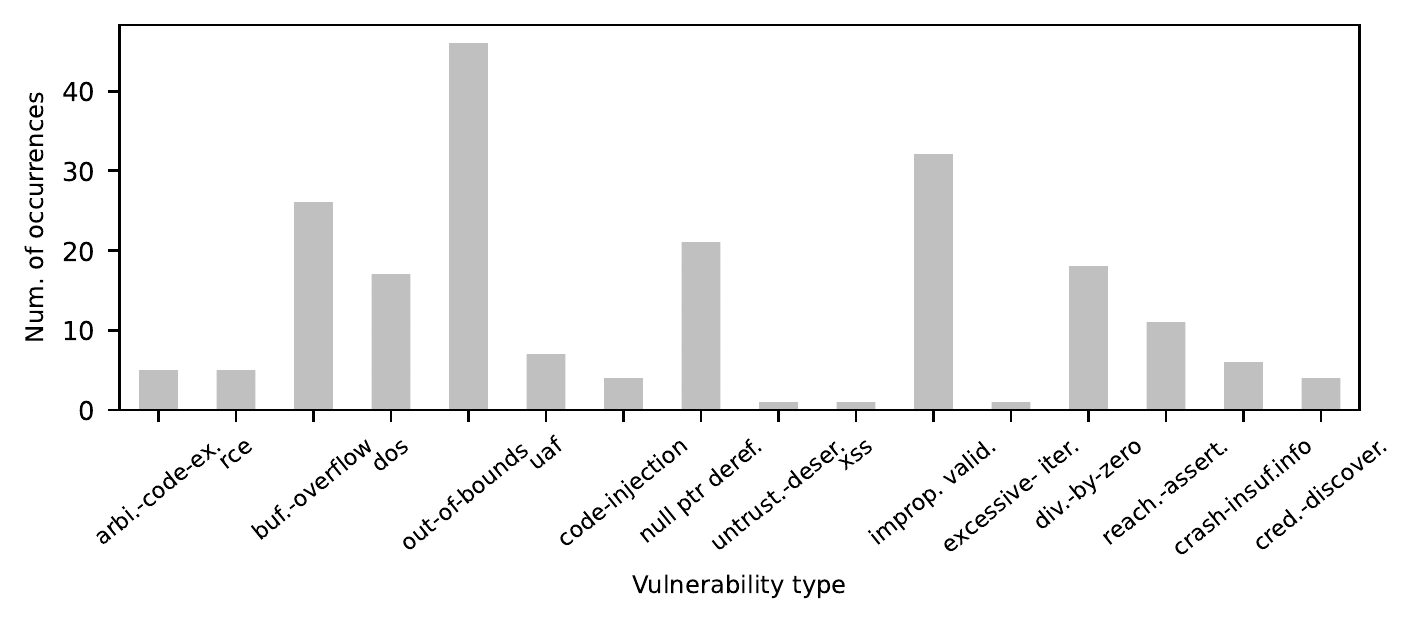}
\caption{Vulnerability importance - GitHub ML repositories 
}
\label{fig:vulnw}
\end{figure}

\subsubsection*{The potential threats from vulnerabilities in the ML repositories}

Most threats found in ML repositories were of two categories: software-level and network-level. Fig.\mbox{~\ref{fig:vulnw}} shows 16 vulnerability types found in the studied ML repositories from the GitHub search and their occurrences. These 16 vulnerabilities can be exploited to cause more than 16 threats on ML systems; since a single vulnerability can cause several damages. Potential threats are grouped as follows: (1) software-level threats include \textit{arbitrary code execution, buffer overflow, denial-of-service (DoS), out-of-bounds (read/write), use-after-free (UAF), code injection, null-pointer dereference, untrusted-deserialization, improper validation, excessive iteration, divide-by-zero, reachable assertion, crash-insufficient information}; and (2) network-level threats include \textit{remote code execution (RCE), cross-site-scripting (XSS),} and \textit{credentials discovery sniffing}. In addition, we want to highlight that software-level DoS are different from network-level DoS that disrupts the normal traffic of a network resource. Software-level DoS can be a memory or crash error (e.g., segmentation fault) that causes disruption of the underlining OS and machine. Threats such as \textit{out-of-bounds}, \textit{buffer overflow}, \textit{improper validation}, \textit{code injection}, \textit{DoS}, and \textit{excessive iteration} were the most prominent in ML repositories extracted from GitHub.

\begin{figure}[h]
\centering
\includegraphics[width=0.49\textwidth]{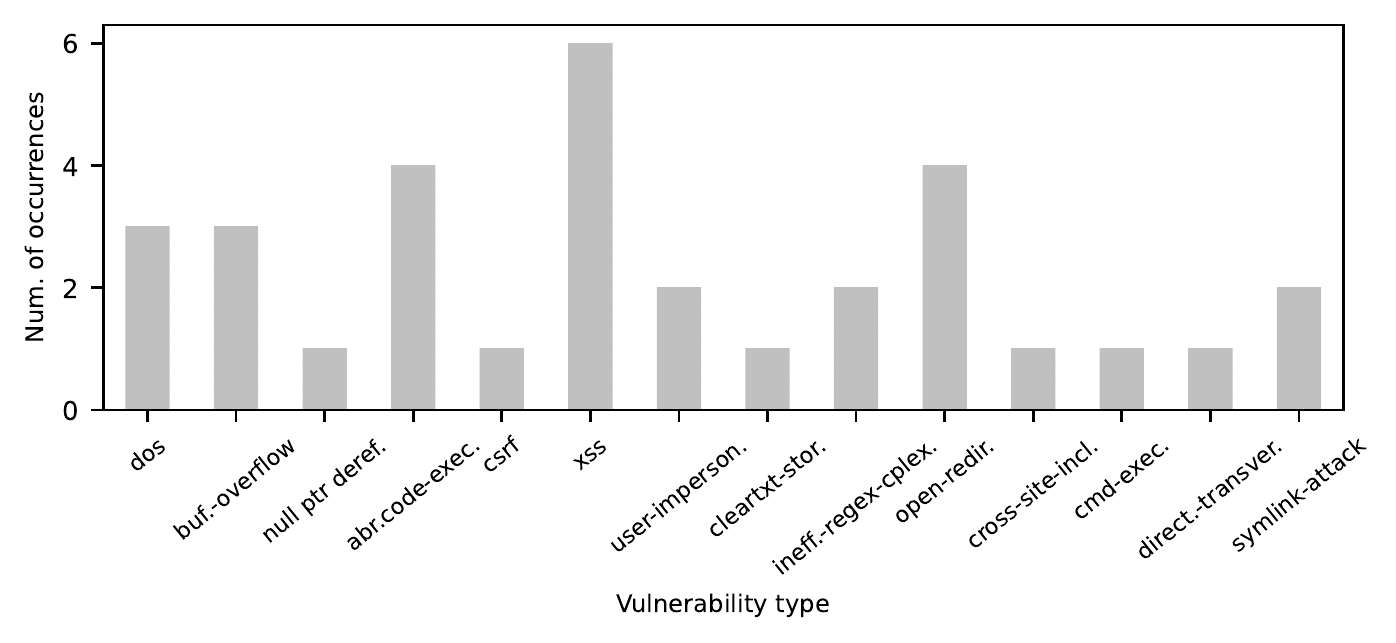}
\caption{Vulnerability importance - PyPA ML repositories 
}
\label{fig:vulnga}
\end{figure}

Fig.\mbox{~\ref{fig:vulnga}} shows 14 additional vulnerability types found in the PyPA database and their occurrences. 
The 14 potential threats are grouped as follows: (1) software-level threats include \textit{DoS, buffer overflow, null-pointer dereference, arbitrary code execution, user impersonation, command execution, cleartext storage, inefficient regex  complexity, symlink}; and (2) network-level threats include \textit{XSS, cross site request forgery (CSRF), cross site inclusion, directory transversal,} and \textit{open redirect}.
The most prominent threats in ML repositories from the PyPA database are \textit{XSS, arbitrary code execution, open redirect, DoS,} and \textit{buffer-overflow}. 

\begin{figure}[h]
\centering
\includegraphics[width=0.49\textwidth]{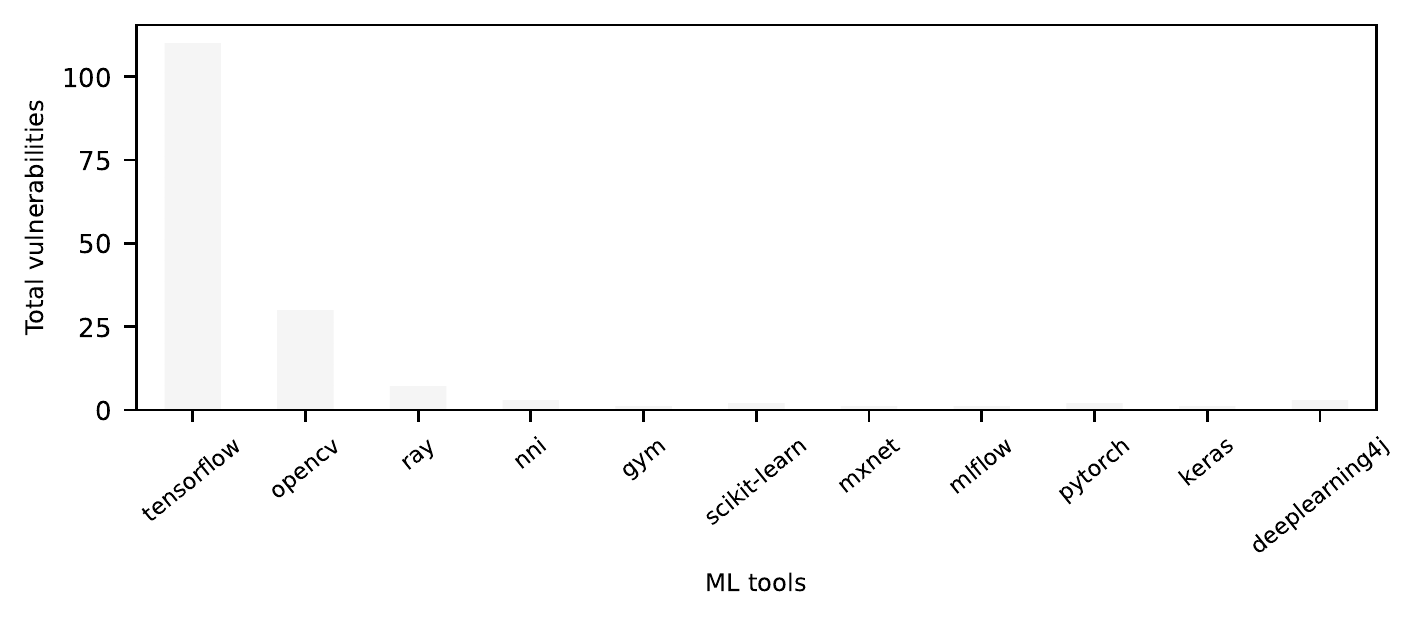}
\caption{Vulnerability importance per tool (from the GitHub search)}
\label{fig:vulntw}
\end{figure}

Globally, buffer-overflow and DoS are prominent in both ML repositories obtained from GitHub and the PyPA database.

\begin{boxblock}{Summary 3}
    \begin{itemize}
      \item The most prominent threats in ML repositories extracted from GitHub are \textit{out-of-bounds}, \textit{buffer overflow}, \textit{improper validation}, \textit{code injection}, and \textit{DoS}
      \item The most prominent threats in ML repositories from the PyPA database are \textit{XSS, arbitrary code execution, open redirect, DoS,} and \textit{buffer-overflow}
      \item The most prominent threats in both ML repositories are \textit{buffer-overflow} and \textit{DoS}
      \item 32 new ML attack scenarios (i.e., 17 techniques, 13 tactics) have been identified and can be used to complete ATLAS case studies for future research
    \end{itemize}
\end{boxblock}


\begin{table}[h]
\caption{\label{depvuln} Dependencies that caused vulnerabilities in repositories}
\resizebox{\columnwidth}{!}{
\begin{tabular}{|l|l|l|l|}
\hline
\textbf{dependency} & \textbf{\begin{tabular}[c]{@{}l@{}}vulner. num. \\ caused\end{tabular}} & \textbf{affected repos.} & \textbf{severity (avg.)} \\ \hline
yaml & 1 & tensorflow & high \\ \hline
sqlite3 & 16 & tensorflow & high \\ \hline
libjpeg-turbo & 4 & tensorflow & medium \\ \hline
giflib & 1 & tensorflow & medium \\ \hline
icu & 1 & tensorflow & high \\ \hline
libjpeg-9c & 1 & tensorflow & high \\ \hline
libpng & 1 & opencv & high \\ \hline
lua54 & 5 & ray & high \\ \hline
pickle & 4 & \begin{tabular}[c]{@{}l@{}}nni, pytorch, \\ pandas, numpy\end{tabular} & critical \\ \hline
requests & 1 & mxnet & high \\ \hline
pyyaml & 1 & nni & high \\ \hline
pillow & 1 & gym & high \\ \hline
log4j & 3 & ray, mlflow & high \\ \hline
numpy116 & 1 & nni & critical \\ \hline
joblib & 1 & scikit-learn & critical \\ \hline
libsvm & 1 & scikit-learn & high \\ \hline
python391 & 1 & pytorch & critical \\ \hline
jquery211 & 1 & keras & medium \\ \hline
curl/libcurl7 & 17 & tensorflow & high \\ \hline
\begin{tabular}[c]{@{}l@{}}commons-\\compress118\end{tabular} & 1 & deeplearnin4j & high \\ \hline
jackson-bind100 & 1 & deeplearnin4j & high \\ \hline
snakeyaml124 & 1 & deeplearnin4j & high \\ \hline
psutil & 1 & tensorflow & high \\ \hline
\end{tabular}
}
\end{table}

\begin{figure*}[h]
\centering
\includegraphics[width=0.99\textwidth]{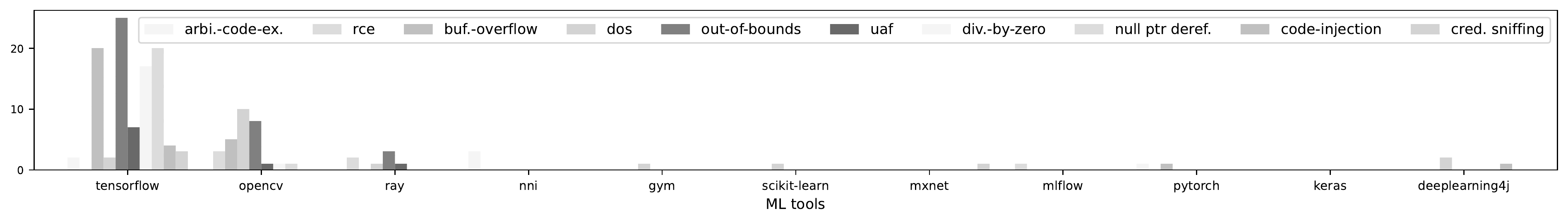}
\caption{Vulnerability distribution across ML repositories (from the GitHub search)}
\label{fig:vulnaccross}
\end{figure*}

\begin{figure*}[h]
\centering
\includegraphics[width=0.99\textwidth]{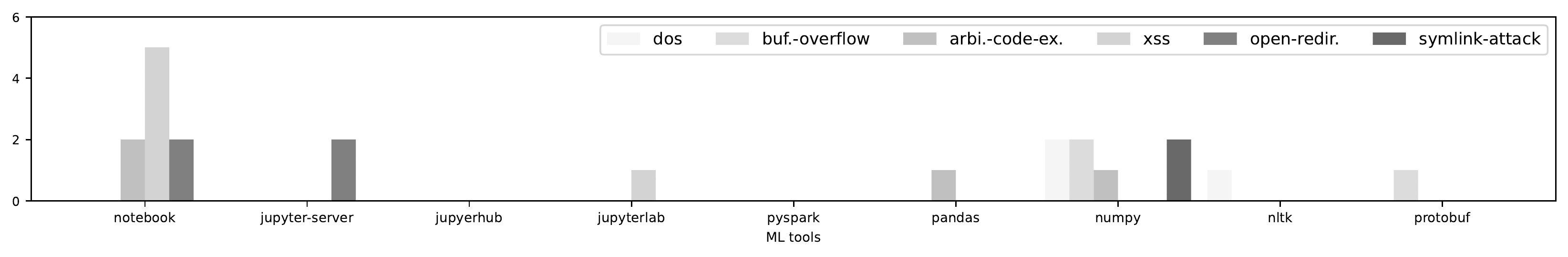}
\caption{Vulnerability distribution across ML repositories (PyPA database)}
\label{fig:vulnaccrossdb}
\end{figure*}

\subsubsection*{The most vulnerable ML repositories as well as the dependencies that cause them}

\begin{figure}[h]
\centering
\includegraphics[width=0.49\textwidth]{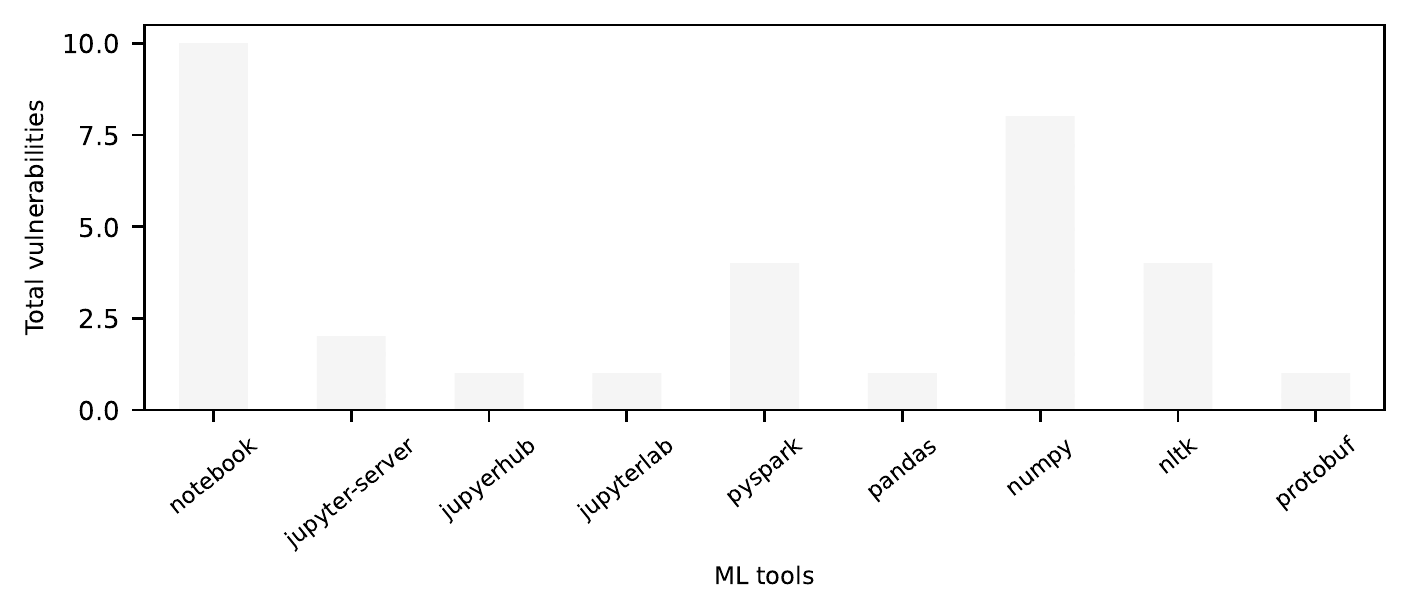}
\caption{Vulnerability importance per tool (PyPA database)}
\label{fig:vulntg}
\end{figure}

Fig.~\ref{fig:vulntw} shows the vulnerability importance per tool from ML repositories from the GitHub search. ML repositories with the highest number of vulnerabilities are \textit{TensorFlow, OpenCV, Ray,} and \textit{NNI}. This can be understood in two manners: (1) these repositories are more vulnerable since their reputation make them a good target for threat actors, and (2) maintaining organisations are more aware of security threats and release vulnerabilities reports as frequently as possible. Other notable repositories such as \textit{Scitkit-Learn, Pytorch, Keras, MLflow, Deeplearning4j} and \textit{Mxnet} have a lower number of vulnerabilities but it does not necessarily mean that they are less vulnerable. This because, it is possible that their maintaining organisations have decided to not 
disclose some vulnerabilities. Fig.~\ref{fig:vulntg} also shows the vulnerability importance per tool from ML repositories in the PyPA database. \textit{Notebook, Numpy, Pyspark,} and \textit{NLTK} have the highest number of vulnerabilities.

Table~\ref{depvuln} shows the dependencies that caused vulnerabilities in the ML repositories by their name, the number of vulnerabilities caused, the affected repositories, and the average severity level according to the CVSS v3/v2 score. Data presented in this table are obtained using the process described 
in Fig.~\ref{nvd-investigation}. 
As shown in Table~\ref{depvuln}, TensorFlow was affected by 5 dependencies with high severity (i.e., \textit{yaml, sqlite3, icu, libjpeg-9c, curl/libcurl7, psutil}) and 2 dependencies with medium severity (i.e., \textit{giflib, libjpeg-turbo}). 

\begin{boxblock}{Summary 4}
    \begin{itemize}
      \item The most vulnerable ML repositories from the GitHub search are \textit{Tensorflow, OpenCV, Ray,}  and \textit{NNI}
      \item The most vulnerable ML repositories from the PyPA database are \textit{Scikit-learn, Pytorch, Keras, MLflow,} and \textit{Mxnet}
      \item The most severe dependencies causing the vulnerabilities are \textit{pickle, joblib, numpy116}, \textit{python3.9.1, log4j, sqlite3, pillow, curl/libcurl7, snakeyaml124}, \textit{commons-compress118}, \textit{jackson-bind100}, \textit{lua54}, \textit{pyyaml, libjpeg-9c,  libsvm, icu,}, \textit{requests}, and \textit{psutil}.
    \end{itemize}
\end{boxblock}

\begin{figure*}[h]
\centering
\includegraphics[width=\textwidth]{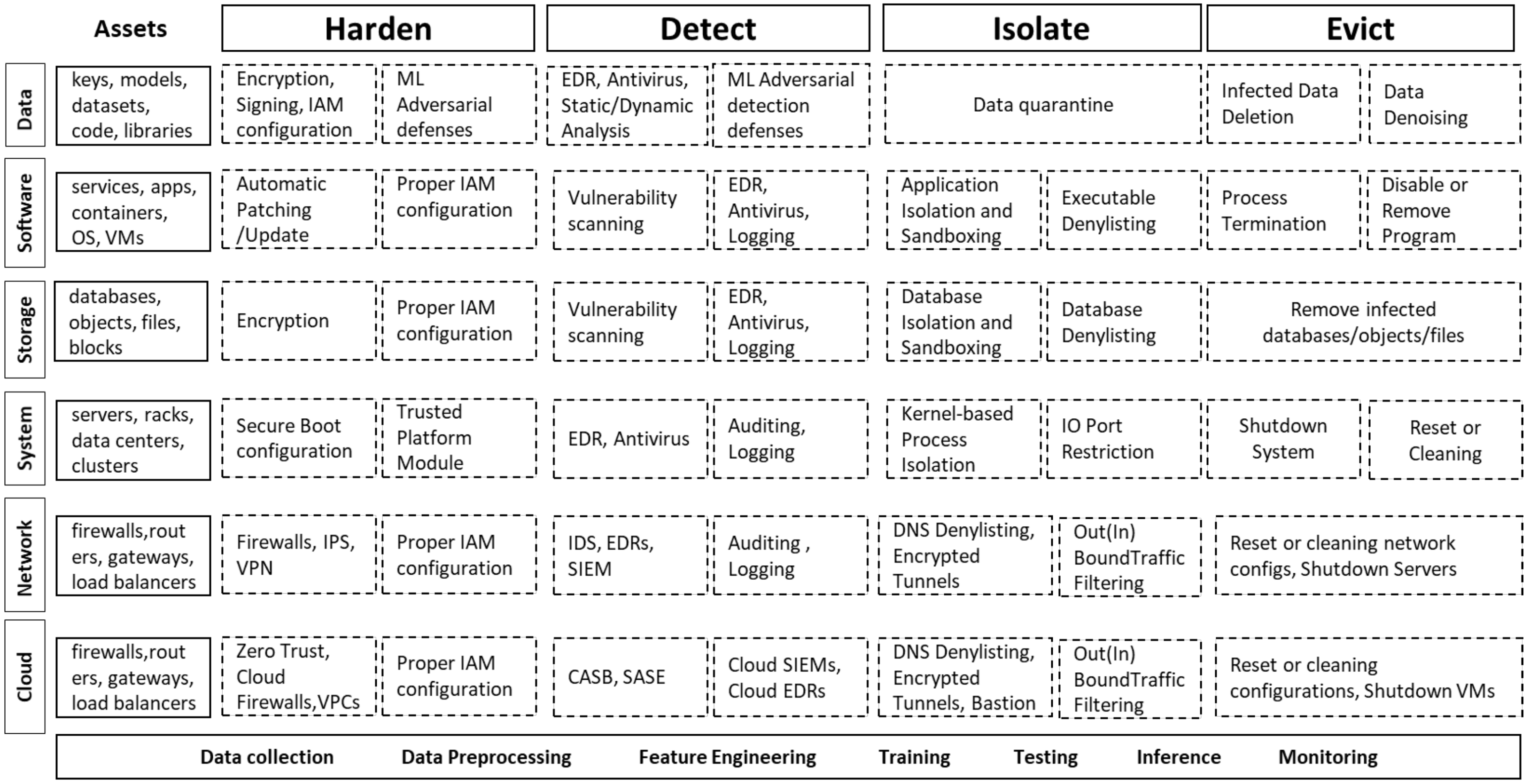}
\caption{ML Threat Mitigation Matrix}
\label{defend_matrix}
\end{figure*}

Dependencies like \textit{pickle, numpy116} (version 1.16.0), \textit{joblib}, and \textit{python391} (version 3.9.1 of python) have a critical severity and affected the following repositories: \textit{Pytorch, NNI, Pandas, Numpy,} and \textit{Scitkit-Learn}. \textit{Deeplearning4j} was affected by 3 dependencies with high severity: \textit{commons-compress118} (Apache version 1.18), \textit{jackson-bind100} (version 1.0.0 beta7), and \textit{snakeyaml124} (version 1.24). ML repositories such as \textit{Ray} and \textit{MLflow} were affected by the \textit{Log4j} dependency which has a high severity.
Other repositories such as OpenCV and Mxnet are respectively affected by libpng and requests dependencies with high severity.

\subsubsection*{The most frequent vulnerabilities in the ML repositories}

After identifying vulnerabilities and the dependencies that caused them, we want to know how they propagated across the studied ML repositories. Fig.~\ref{fig:vulnaccross} shows the vulnerability distribution in ML repositories 
obtained from the GitHub search. The most frequent vulnerability is \textit{DoS}. It occurs 
in 6 repositories: \textit{Tensorflow, OpenCV, Deeplearning4j, Ray, Scikit-Learn,} and \textit{Gym}. The next ones are \textit{arbitrary code execution, RCE, UAF,} and \textit{buffer overflow}. The \textit{arbritrary code execution} vulnerability is found in 3 repositories: \textit{TensorFlow, NNI,} and \textit{Pytorch}. \textit{RCE} spreads on 3 repositories: \textit{OpenCV, Ray,} and \textit{MLflow}. \textit{UAF} occurs in 3 repositories: \textit{TensorFlow, OpenCV,} and \textit{Ray}. \textit{Buffer overflow} is found in 3 repositories: \textit{TensorFlow, OpenCV,} and \textit{Pytorch}. Other vulnerabilities such as \textit{out-of-bounds, code-injection, null-pointer dereference,} and \textit{credential sniffing} were respectively detected in 2 repositories (i.e., \textit{OpenCV} and \textit{Ray}, \textit{TensorFlow} and \textit{OpenCV}, \textit{TensorFlow} and \textit{Deeplearning4j}, \textit{TensorFlow} and \textit{Mxnet}). 

Fig.~\ref{fig:vulnaccrossdb} shows the vulnerability distribution in ML repositories 
from the PyPA database. The most frequent vulnerability is 
\textit{arbitrary code execution}. It occurs in 3 repositories: \textit{Notebook, Pandas,} and \textit{Numpy}. Then, other vulnerabilities such as \textit{XSS, DoS, buffer overflow,} and \textit{open redirect} are found each in 2 repositories; respectively, \textit{Notebook} and \textit{Jupyterlab}, \textit{Numpy} and \textit{NLTK}, \textit{Numpy} and \textit{Protobuf}, \textit{Notebook} and \textit{Jupyter-server}.

\begin{boxblock}{Summary 5}
    \begin{itemize}
      \item The most frequent vulnerabilities in ML repositories from the GitHub search are \textit{DoS, arbitrary code execution, RCE, UAF,} and \textit{buffer overflow}
      \item The most frequent vulnerabilities in ML repositories from the PyPA database are \textit{arbitrary code execution, XSS, DoS, buffer overflow,} and \textit{open redirect}
      \item The most frequent vulnerabilities in both ML repositories are \textit{DoS, buffer overflow,} and \textit{arbitrary code execution}
    \end{itemize}
\end{boxblock}


%
%

\section{Mitigation of vulnerabilities and threats}~\label{mitigation}
The previous sections identified and characterized 
 ML threats, as well as their impact on ML components (models, phases, tools). In order to answer to RQ4, we provide proactive and reactive countermeasures that could be used to mitigate these ML threats. Based on security guidance from MITRE ATT\&CK~\cite{mitre_attack_mitigation}, MITRE D3FEND~\cite{mitre_defend}, NIST security guidelines~\cite{mccarthy1800identity, hu2020general, souppaya2013guide, scarfone2008technical, scarfone2008sp, karygiannis2002wireless, cooper2018security}, and the Cloud Security Alliance~\cite{csa_report}, we propose a mitigation matrix to help harden ML assets, detect vulnerability and threats on ML assets, isolate and evict infected ML assets from data level to cloud level during the ML lifecycle (see Fig.~\ref{defend_matrix}). In the following, some mitigations are described and the documentation in progress can be found in~\cite{tech-report-v1}.

At \textit{data} level, hardening techniques include adversarial defenses~\cite{tabassi2019taxonomy, arp2022and, tramer2020adaptive}; and tools such as ART, cleverhans, foolbox. Code signing certificates~\cite{cooper2018security} can be used to ensure authenticity and integrity of the data sources. Data protection~\cite{scarfone2008sp, csa_report} can be enforced \textit{in transit} by using TLS encryption to secure ML API calls, \textit{at rest} by using AES-based keys for protecting ML assets during ML phases, and \textit{in use} by dynamic analysis of data. Proper identity and access management (IAM) based on the least privilege~\cite{mccarthy1800identity, marshall2010security, gcp_foundation_guide} can be used to limit unintended access on ML artifacts such as  training data, ML models, model parameters, and model results produced/used during ML phases. To detect adversarial examples,  techniques such as Introspection~\cite{aigrain2019detecting}, Feature Squeezing~\cite{xu2017feature}, and SafetyNet\cite{lu2017safetynet} can be used. When an attack is identified, infected training data and ML models can be stored in isolated databases/cloud environments~\cite{mundada2011silverline}. Then, training data and ML models can be deleted or sanitized using denoisers~\cite{xie2019feature, liao2018defense, gu2014towards}.

At \textit{software} level, regular vulnerability scanning of ML libraries, ML pipeline and model source code~\cite{scarfone2008technical} can be used to mitigate vulnerabilities in ML systems using tools such as GitHub code scanning, Tsunami, OSS-fuzz, and SonaQube. IAM based on least privilege~\cite{mccarthy1800identity} can be enforced on running processes, applications, containers, and virtual machines to restrict accesses on premises and off premises. At \textit{database} level, proper IAM configuration~\cite{mccarthy1800identity, marshall2010security, gcp_foundation_guide} can be used to limit database accesses. Database backup must be regularly done and kept separate from the corporate network to prevent compromise~\cite{chandramouli2020security}. At \textit{system} level, proper OS hardening~\cite{souppaya2013guide} such as automatic system updates, Secure Boot enabling and configuration, and limitation of access permission can be enforced to secure the ML system endpoints. Regular patch management~\cite{souppaya2013guide} with up-to-date ML libraries can be enforced to mitigate the vulnerabilities in ML systems. For threat detection, ML infrastructures can use endpoint detection and response (EDRs) or antivirus (e.g., Microsoft Defender, Wazuh), security information and event management systems (SIEMs), audit and logging. 

At \textit{network} level, network access control lists~\cite{souppaya2016guide01, hu2020general,frankel2005guide} can be applied to virtual private clouds (VPCs) to control traffic access for virtual machine instances of the ML infrastructure. Firewalls~\cite{wack2002guidelines} and Intrusion Prevention Systems (IPS)~\cite{scarfone2007guide} can be installed as first line of defense to analyze and take immediate actions when a malicious traffic is observed. Encrypted tunnels (e.g., IPSec VPN~\cite{frankel2005guide}) can be used to ensure end-to-end security of remote accesses in the corporate network and external networks (e.g., partners, customers). At \textit{cloud} level, security mechanisms~\cite{csa_report, jansen2011guidelines} include IAM with policies based on least privilege, role-based permissions, and multi-factor authentication (MFA); enforcing cloud security policies using solutions such as cloud access security broker (CASB), Zero trust~\cite{rose2020zero, zero-trust-paper}, and secure access service edge (SASE); endpoint protection using EDRs (e.g., Falcon Insight, Cortex XDR, Microsoft Defender); network prevention and detection systems (e.g., Snort, Zeek); cloud-based SIEMs (e.g., Splunk Cloud, Azure Sentinel); setting of bastion/transit virtual networks for more-flexible hybrid clouds; auditing and logging.


\section{Threats to validity}~\label{threat2valid}

ATLAS is a new database less mature than ATT\&CK (e.g., it does not yet have specific mitigation definitions for ML threats, there is no tool support). In addition, The AI Incident database is used to collect new threat feeds but does not provide clear information of the TTPs and models used for the attacks; thus limiting the scope of the study. Moreover, mining vulnerabilities in large ML repositories is a challenging task that requires automation and can be slow. Some vulnerabilities can not be found because some maintainers did not report them through issues or security advisories. Moreover, searching for hidden vulnerabilities from references in issues (i.e., mailing lists, discussion threads, websites) is time consuming and some vulnerabilities could be missed. Precisely, issue comments and references often contain CVE codes with typo errors (e.g., CVE-2018 without the number) or non-ascii parts (e.g., CVE-2017\textbackslash x01124) that could be missed using regular expressions. Nevertheless, we spent effort to systematically collect threat/vulnerability data while ensuring they are reliable. We also proposed mitigations for each specific ML phase and asset to ensure threat prevention. 


\section{Conclusion}~\label{conclusion}

In this work, we have studied ML threat behaviors and analyzed the impact of these threats on ML components using ATLAS and AI Incident databases; and ML repositories from GitHub and the PyPA database. 
Results show that CNNs were one of the most targeted in attack scenarios.
Our examinations of vulnerabilities and attacks reveal that testing,
inference, training are the most targeted
ML phases by threat actors. Threats like buffer-overflow and DoS were the most frequent threats in the studied ML repositories. 
ML repositories such as TensorFlow, OpenCV, and Notebook
have the largest vulnerability prominence and the most severe dependencies causing them include \textit{pickle, joblib, numpy116, python3.9.1,} and \textit{log4j}. 32 new ML attack scenarios (17 techniques, 13 tactics) have been identified and can be used
to complete ATLAS case studies for future research. To help mitigate these threats, we have proposed an ML threat mitigation matrix to help defend against real-world 
threats targeting ML products and ML cloud infrastructures. In the future, we plan to implement a ML threat assessment tool 
to continuously ingest various vulnerability feeds, TTP definitions from several ML threat databases, and generate attack matrices and analytic graphs in real-time using Natural Language Processing techniques. We also plan to contribute to the development of tools to support the ATLAS framework.

\section*{Acknowledgment}

This work is partly funded by the Fonds de Recherche du Québec (FRQ), Natural Sciences and Engineering Research Council of Canada (NSERC), Canadian Institute for Advanced Research (CIFAR), and Mathematics of Information Technology and Complex Systems (MITACS).



\bibliographystyle{IEEEtran}
\balance
\bibliography{bibliography}
\end{document}